\documentclass[conference]{IEEEtran}
\usepackage[boxruled]{algorithm2e}
\usepackage{cite}
\usepackage{graphicx}
\usepackage{psfrag}
\usepackage{subfigure}
\usepackage{url}
\usepackage{amsmath}
\usepackage{array}
\usepackage{amssymb}
\usepackage{amsfonts}
\usepackage{graphicx}
\usepackage{epstopdf}

\newtheorem{lemma}{Lemma}

\newtheorem{definition}{Definition}
\newtheorem{theorem}{Theorem}
\newtheorem{example}{Example}
\newtheorem{note}{Note}

\title{Channel Quantization for Physical Layer Network-Coded Two-Way Relaying}
\begin{document}


\author{
\authorblockN{Vijayvaradharaj T Muralidharan, Vishnu Namboodiri and B. Sundar Rajan }
\authorblockA{Dept. of ECE, IISc, Bangalore 560012, India, Email: {$\lbrace$ tmvijay, vishnukk, bsrajan$\rbrace$} @ece.iisc.ernet.in
}
}

\maketitle
\thispagestyle{empty}	
\begin{abstract}
 The design of modulation schemes for the physical layer network-coded two way relaying scenario is considered with the protocol which employs two phases: Multiple access (MA) Phase and Broadcast (BC) phase. It was observed by Koike-Akino et al. that adaptively changing the network coding map used at the relay according to the channel conditions greatly reduces the impact of multiple access interference which occurs at the relay during the MA phase. In other words, the set of all possible channel realizations (the complex plane) is quantized into a finite number of regions, with a specific network coding map giving the best performance in a particular region. We highlight the issues associated with the scheme proposed by Koike-Akino et al. and propose a scheme which solves these issues. We obtain a quantization of the  set of all possible channel realizations analytically for the case when $M$-PSK (for $M$ any power of $2$) is the signal set used during the MA phase. It is shown that the complex plane can be classified into two regions: a region in which any network coding map which satisfies the so called exclusive law gives the same best performance and a region in which the choice of the network coding map affects the performance, which is further quantized based on the choice of the network coding map which optimizes the performance. The quantization thus obtained analytically, leads to the same as the one obtained using computer search for 4-PSK signal set by Koike-Akino et al., when specialized for $M=4.$ 
\end{abstract}

\section{BACKGROUND AND PRELIMINARIES}

\begin{figure}[htbp]
\centering
\subfigure[MA Phase]{
\includegraphics[totalheight=1in,width=2in]{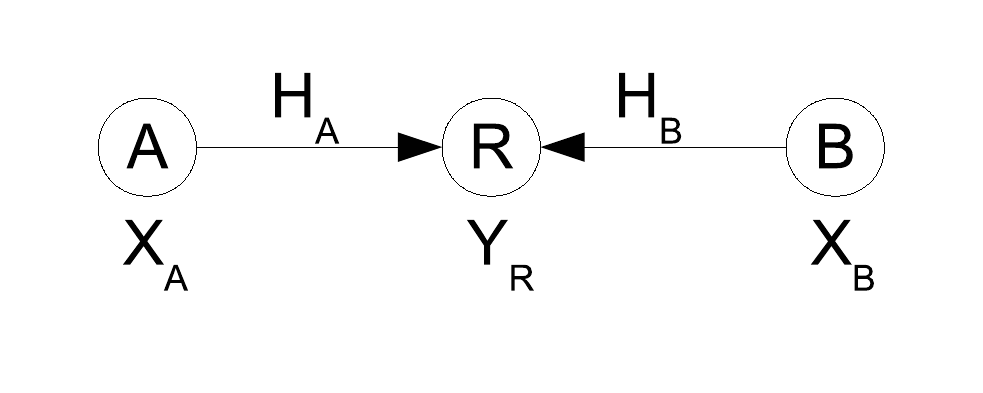}
\label{fig:phase1}	
}

\subfigure[BC Phase]{
\includegraphics[totalheight=1in,width=2in]{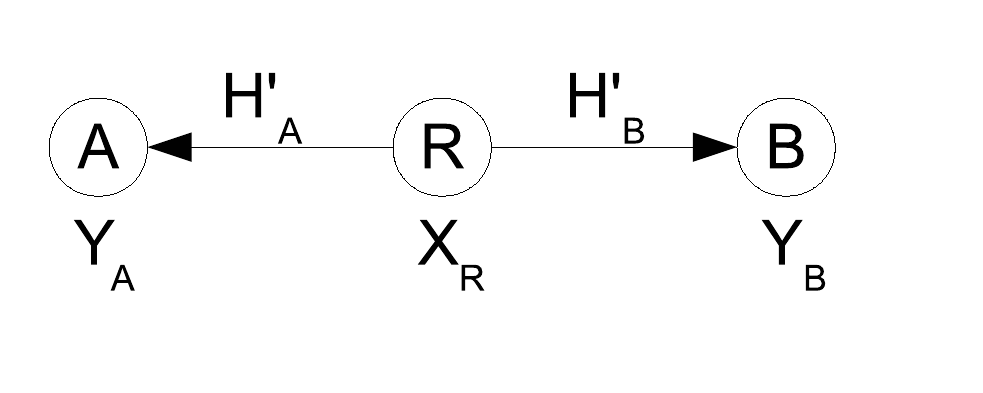}
\label{fig:phase2}	
}
\caption{The Two Way Relay Channel}
\label{relay_channel}
\end{figure}

We consider the two-way wireless relaying scenario shown in Fig. \ref{relay_channel}, where bi-directional data transfer takes place between the nodes A and B with the help of the relay R. It is assumed that all the three nodes operate in half-duplex mode, i.e., they cannot transmit and receive simultaneously in the same frequency band. The relaying protocol consists of the following two phases: the \textit{multiple access} (MA) phase, during which A and B simultaneously transmit to R and the \textit{broadcast} (BC) phase during which R transmits to A and B. Network coding is employed at R in such a way that A (B) can decode the message of B (A), given that A (B) knows its own message.


 The concept of physical layer network coding has attracted a lot of attention in recent times. The idea of physical layer network coding for the two way relay channel was first introduced in \cite{ZhLiLa}, where the multiple access interference occurring at the relay was exploited so that the communication between the end nodes can be done using a two stage protocol. Information theoretic studies for the physical layer network coding scenario were reported in \cite{KiMiTa},\cite{PoYo}. The design principles governing the choice of modulation schemes to be used at the nodes for uncoded transmission were studied in \cite{KoPoTa}. An extension for the case when the nodes use convolutional codes was done in \cite{KoPoTa_conv}. A multi-level coding scheme for the two-way relaying scenario was proposed in \cite{HeNa}.

It was observed in \cite{KoPoTa} that for uncoded transmission, the network coding map used at the relay needs to be changed adaptively according to the channel fade coefficient, in order to minimize the impact of the multiple access interference. In other words, the set of all possible channel realizations is quantized into a finite number of regions, with a specific network coding map giving the best performance in a particular region. In \cite{KoPoTa}, a quantization was obtained using computer search for 4-PSK signal set.  In this paper, we obtain such a quantization analytically for any $M$-PSK signal set, where $M=2^\lambda$, for some positive integer $\lambda$.
\subsection{Signal Model}
\vspace{-0.1 cm}
\subsubsection*{Multiple Access (MA) Phase}

Let $\mathcal{S}$ denote the $M$-PSK constellation used at A and B, where $M$ is of the form $2^\lambda$, $\lambda$ being a positive integer. Assume that A (B) wants to transmit an $\lambda$-bit binary tuple to B (A). Let $\mu: \mathbb{F}_{2^\lambda} \rightarrow \mathcal{S}$ denote the mapping from bits to complex symbols used at A and B. Let $x_A= \mu(s_A)$, $x_B=\mu(s_B)$ $\in \mathcal{S}$ denote the complex symbols transmitted by A and B respectively, where $s_A,s_B \in \mathbb{F}_{2^\lambda}$. The received signal at $R$ is given by,

\begin{align}
\nonumber
Y_R=H_{A} x_A + H_{B} x_B +Z_R,
\end{align}
where $H_{A}$ and $H_{B}$ are the fading coefficients associated with the A-R and B-R links respectively. The additive noise $Z_R$ is assumed to be $\mathcal{CN}(0,\sigma^2)$, where $\mathcal{CN}(0,\sigma^2)$ denotes the circularly symmetric complex Gaussian random variable with variance $\sigma ^2$. We assume a block fading scenario, with the ratio $ H_{B}/H_{A}$ denoted as $z=\gamma e^{j \theta}$ referred to as the fading state, where $\gamma \in \mathbb{R}^+$ and $-\pi \leq \theta < \pi$. Also, it is assumed that $z$ is distributed according to a continuous probability distribution.   
 
 Let $\mathcal{S}_{R}(\gamma,\theta)$ denote the effective constellation at the relay during the MA Phase, i.e., 

{\footnotesize
\begin{align} 
\nonumber
 \mathcal{S}_{R}(\gamma,\theta)=\left\lbrace s_i+\gamma e^{j \theta} s_j \vert s_i,s_j \in \mathcal{S}\right \rbrace.
 \end{align}
}
  
Let $d_{min}(\gamma e^{j\theta})$ denote the minimum distance between the points in the constellation $\mathcal{S}_{R}(\gamma,\theta)$, i.e.,

{\footnotesize
\begin{align}
\label{eqn_dmin} 
d_{min}(\gamma e^{j\theta})=\hspace{-0.5 cm}\min_{\substack {{(x_A,x_B),(x'_A,x'_B)}\\{ \in \mathcal{S} \times \mathcal{S}} \\ {(x_A,x_B) \neq (x'_A,x'_B)}}}\hspace{-0.5 cm}\vert \left(x_A-x'_A\right)+\gamma e^{j \theta} \left(x_B-x'_B\right)\vert.
\end{align}
}
 
 From \eqref{eqn_dmin}, it is clear that there exists values of $\gamma e^{j \theta}$ for which $d_{min}(\gamma e^{j\theta})=0$. Let $\mathcal{H}=\lbrace \gamma e^{j\theta} \in \mathbb{C} \vert d_{min}(\gamma,\theta)=0 \rbrace$. The elements of $\mathcal{H}$ are said to be the singular fade states. 

\begin{definition}
 A fade state $\gamma e^{j \theta}$ is said to be a singular fade state, if the cardinality of the signal set $\mathcal{S}_{R}(\gamma, \theta)$ is less than $M^2$.
 \end{definition}
 
 For example, consider the case when symmetric 4-PSK signal set used at the nodes A and B, i.e., $\mathcal{S}=\lbrace (\pm 1 \pm j)/\sqrt{2} \rbrace$. For $\gamma e^{j \theta}=(1+j)/2$, $d_{min}(\gamma e^{j \theta})=0$, since,

{\footnotesize
\begin{align*} 
 \left\vert \left( \dfrac{1+j}{\sqrt{2}}-\dfrac{1-j}{\sqrt{2}} \right) + \dfrac{(1+j)}{2} \left( \dfrac{-1-j}{\sqrt{2}} - \dfrac{1+j}{\sqrt{2}} \right)\right\vert=0.
 \end{align*}
 }
 
Alternatively, when $\gamma e^{j \theta}=(1+j)/2$, the constellation $\mathcal{S}_{R}(\gamma,\theta)$ has only 12 ($<$16) points. 
 Hence $\gamma e^{j \theta}=(1+j)/2$ is a singular fade state for the case when 4-PSK signal set is used at A and B.
 
 
 
 Let $(\hat{x}_A,\hat{x}_B) \in \mathcal{S} \times \mathcal{S}$ denote the Maximum Likelihood (ML) estimate of $({x}_A,{x}_B)$ based on the received complex number $Y_{R}$, i.e.,
 
 {\footnotesize
 \begin{align}
 (\hat{x}_A,\hat{x}_B)=\arg\min_{({x}'_A,{x}'_B) \in \mathcal{S} \times \mathcal{S}} \vert Y_R-H_{A}{x}'_A-H_{B}{x}'_B\vert.
 \end{align}
 }

\subsubsection*{Broadcast (BC) Phase}

Depending on the value of $\gamma e^{j \theta}$, R chooses a map $\mathcal{M}^{\gamma,\theta}:\mathcal{S} \times \mathcal{S} \rightarrow \mathcal{S}'$, where $\mathcal{S}'$ is the signal set used by R during $BC$ phase. The elements in $\mathcal{S} \times \mathcal{S}$ which are mapped on to the same complex number in $\mathcal{S}'$ by the map $\mathcal{M}^{\gamma,\theta}$ are said to form a cluster. Let $\lbrace \mathcal{L}_1, \mathcal{L}_2,...,\mathcal{L}_l\rbrace$ denote the set of all such clusters. The formation of clusters is called clustering, denoted by $\mathcal{C}^{\gamma,\theta}$. 
 

The received signals at A and B during the BC phase are respectively given by,

\begin{align}
Y_A=H'_{A} X_R + Z_A,\;Y_B=H'_{B} X_R + Z_B,
\end{align}
where $X_R=\mathcal{M}^{\gamma,\theta}(\hat{x}_A,\hat{x}_B) \in \mathcal{S'}$ is the complex number transmitted by R. The fading coefficients corresponding to the R-A and R-B links are denoted by $H'_{A}$ and $H'_{B}$ respectively and the additive noises $Z_A$ and $Z_B$ are $\mathcal{CN}(0,\sigma ^2$).

In order to ensure that A (B) is able to decode B's (A's) message, the clustering $\mathcal{C}^{\gamma,\theta}$ should satisfy the exclusive law \cite{KoPoTa}, i.e.,

{\footnotesize
\begin{align}
\nonumber
\mathcal{M}^{\gamma,\theta}(x_A,x_B) \neq \mathcal{M}^{\gamma,\theta}(x'_A,x_B), \; \mathrm{where} \;x_A \neq x'_A \; \mathrm{,} \;x_B \in  \mathcal{S},\\
\label{ex_law}
\mathcal{M}^{\gamma,\theta}(x_A,x_B) \neq \mathcal{M}^{\gamma,\theta}(s_A,s'_B), \; \mathrm{where} \;x_B \neq x'_B \; \mathrm{,} \;x_A \in \mathcal{S}.
\end{align}
}

From an information theoretic perspective, the mapping $\mathcal{M}^{\gamma,\theta}$ needs to satisfy the exclusive law for the reason outlined below. Consider the ideal situation where the additive noises at the nodes are zero. It is assumed that the fading state $\gamma e^{j \theta} \notin \mathcal{H}$. The assumption is required since R can decode unambiguously to an element in $\mathcal{S} \times \mathcal{S}$ only if $\gamma e^{j \theta}$ is not a singular fade state and is justified since $\gamma e^{j \theta}$ takes values from a continuous probability distribution and the cardinality of $\mathcal{H}$ is finite. During the MA Phase, assume the relay jointly decodes correctly to the pair $(x_A,x_B)$ and transmits $X_R=\mathcal{M}^{\gamma,\theta}(x_A,x_B)$ during the BC Phase. The received complex symbols at A and B are respectively $Y_A=H'_A X_R$ and $Y_B=H'_B X_R$. At node A, the amount of uncertainty about $x_B$ which gets resolved after observing $Y_A$, $I(x_B;Y_A/x_A)=H(x_B\vert x_A)-H(x_B\vert Y_A,x_A)=H(x_B)-H(x_B \vert X_R, x_A)$ (since $x_B$ and $x_A$ are independent). Since $H(x_B \vert X_R, x_A)=0$ if and only if the mapping $\mathcal{M}^{\gamma,\theta}$ satisfies the exclusive law, the amount of uncertainty about $x_B$ ($x_A$) which gets resolved at A (B) is maximized if and only if the clustering satisfies the exclusive law. 

\begin{definition}
The cluster distance between a pair of clusters $\mathcal{L}_i$ and $\mathcal{L}_j$ is the minimum among all the distances calculated between the points $x_A+\gamma e^{j\theta} x_B ,x'_A+\gamma e^{j\theta} x'_B \in \mathcal{S}_R(\gamma,\theta)$ which satisfy  the conditions $(x_A,x_B) \in \mathcal{L}_i$ and $(x'_A,x'_B) \in \mathcal{L}_j$.
\end{definition}
\begin{definition}
The \textit{minimum cluster distance} of the clustering $\mathcal{C}^{\gamma,\theta}$ is the minimum among all the cluster distances, i.e.,

{\footnotesize
\begin{align}
\nonumber
d_{min}(\mathcal{C}^{\gamma,\theta})=\hspace{-0.8 cm}\min_{\substack {{(x_A,x_B),(x'_A,x'_B)}\\{ \in \mathcal{S}\times\mathcal{S},} \\ {\mathcal{M}^{\gamma,\theta}(x_A,x_B) \neq \mathcal{M}^{\gamma,\theta}(x'_A,x'_B)}}}\hspace{-0.8 cm}\vert \left( x_A-x'_A\right)+\gamma e^{j \theta} \left(x_B-x'_B\right)\vert.
\end{align}
}

\end{definition}
The minimum cluster distance determines the performance during the MA phase of relaying. The performance during the BC phase is determined by the minimum distance of the signal set $\mathcal{S}'$. Throughout, we restrict ourselves to optimizing the performance during the MA phase. For values of $\gamma e^{j \theta}$ close to the singular fade states, the value of $d_{min}(\gamma e^{j\theta})$ is greatly reduced. Hence for each singular fade state, a clustering needs to be chosen such that the minimum cluster distance at the singular fade state is non-zero and is also maximized.

A clustering $\mathcal{C}^{\lbrace h \rbrace}$ is said to remove a singular fade state $ h \in \mathcal{H}$, if the minimum cluster distance of the clustering $\mathcal{C}^{\lbrace h \rbrace}$ for $\gamma e^{j \theta}=h$ is greater than zero. 
Let $\mathcal{C}_{\mathcal{H}}=\left\lbrace \mathcal{C}^{\lbrace h\rbrace} : h \in \mathcal{H} \right\rbrace$ denote the set of all such clusterings. For examples of clusterings, see \cite{KoPoTa}, \cite{LS}. In \cite{KoPoTa}, the clusterings were obtained using a computer search algorithm, whereas in \cite{LS}, it is obtained analytically using the mathematical structure called Latin Square. Let $d_{min}({\mathcal{C}^{\lbrace h\rbrace}},\gamma,\theta)$ denote the minimum cluster distance of the clustering $\mathcal{C}^{\lbrace h\rbrace}$ evaluated at $\gamma e^{j\theta}$. For $\gamma e^{j \theta} \notin \mathcal{H}$, the clustering $\mathcal{C}^{\gamma,\theta}$ is chosen to be $\mathcal{C}^{\lbrace h\rbrace}$, which satisfies $d_{min}({\mathcal{C}^{\lbrace h\rbrace}},\gamma,\theta) \geq d_{min}({\mathcal{C}^{\lbrace h' \rbrace}},\gamma,\theta), \forall h \neq h' \in \mathcal{H}$.


\begin{note}
The clusterings which belong to the set $\mathcal{C}_{\mathcal{H}}$ need not be distinct, since a single clustering can remove more than one singular fade state.
\end{note}

\subsection{Issues with Koike-Akino Popovski Tarokh's approach \cite{KoPoTa}}
It is assumed that the channel state information is not available at the transmitting nodes A and B during the MA phase. A block fading scenario is assumed and the clustering used by the relay is indicated to A and B by using overhead bits. 

The procedure suggested in \cite{KoPoTa} to obtain the set of all clusterings, was using a computer algorithm (closest neighbour clustering algorithm), which involved varying the fade state values over the entire complex plane, i.e., $0 \leq  \gamma < \infty$, $0 \leq \theta < 2\pi$ and finding the clustering for each value of channel realization. The total number of network codes which would result is known only after the algorithm is run for all possible realizations $\gamma e^{j \theta}$ which is uncountably infinite and hence the number of overhead bits required is not known beforehand. Moreover, performing such an exhaustive search is extremely difficult in practice, especially when the cardinality of the signal set $M$ is large.

The implementation complexity of the scheme suggested in \cite{KoPoTa} is extremely high. It appears that, for each realization of the singular fade state, the closest neighbour clustering algorithm \cite{KoPoTa} needs to be run at R to find the clustering. In contrast, we provide a simple criterion (Section IV A) based on which a clustering from the set $\mathcal{C}_{\mathcal{H}}$ is chosen depending on the value of $\gamma e^{j \theta}$. In this way, the set of all values of $\gamma e^{j\theta}$ (the complex plane) is quantized into different regions, with a clustering from the set $\mathcal{C}_{\mathcal{H}}$ used in a particular region. 


In the closest neighbour clustering algorithm suggested in \cite{KoPoTa}, the network coding map is obtained by considering the entire distance profile. The disadvantages of such an approach are two-fold. 

\begin{itemize}
\item
Considering the entire distance profile, instead of the minimum cluster distance alone which contributes dominantly to the error probability, results in an extremely large number of network coding maps. For example, for 8-PSK signal set, the closest neighbour clustering algorithm results in more than 5000 maps.
\item
 The closest neighbour clustering algorithm tries to optimize the entire distance profile, even after clustering signal points which contribute the minimum distance. As a result, for several channel conditions, the number of clusters in the clustering obtained is greater than the number of clusters in the clustering obtained by taking the minimum distance alone into consideration. This results in a degradation in performance during the BC Phase, since the relay uses a signal set with cardinality equal to the number of clusters. 
 \end{itemize}

In \cite{KoPoTa}, to overcome the two problems mentioned above, another algorithm is proposed, in which for a given $\gamma e^{j \theta}$, an exhaustive search is performed among all the network coding maps obtained using the closest neighbour clustering algorithm and a map with minimum number of clusters is chosen. The difficulties associated with the implementation of the closest neighbour clustering algorithm carry over to the implementation of this algorithm as well.

To avoid all these problems we suggest a scheme, which is based on the removal of all the singular fade states. Since the number of singular fade states is finite (the exact number of singular fade states and their location in the complex plane are discussed in Section II), the total number of network coding maps used is upper bounded by the number of singular fade states. In fact, the total number of network coding maps required is shown to be lesser than the total number of singular fade states in Section VI in \cite{LS}. In other words, the total number of network coding maps required is known exactly, which determines the number of overhead bits required. It is shown in section III in \cite{LS} that the problem of obtaining clusterings which remove all the singular fade states reduces to completing a finite number of partially filled Latin Squares, which totally avoids the problem of performing exhaustive search for an uncountably infinite number of values.


The contributions and organization of the paper are as follows:
\begin{itemize}
\item
It is shown that the singular fade states lie on $M^2/4-M/2+1$ circles, with $M$ singular fade states lying on each circle, when $M$-PSK signal set is used at A and B (Section II).
\item
It is shown analytically that the $\gamma e^{j \theta}$ (complex) plane can be classified into two regions: a region called the singularity-free region containing no singular fade states in which any clustering satisfying the exclusive law gives the same minimum cluster distance and the other region called the singularity region in which the choice of the clustering impacts the minimum cluster distance (Section III A).
\item
For the case when $M$-PSK signal set is used at A and B, the explicit boundaries of the singularity-free region are obtained (Section III B).
\item
A quantization of the singularity region for $M$-PSK signal set is obtained analytically, with each one of the partitioned regions containing one singular fade state (Section IV A and IV B).
\item
As an example of the channel quantization obtained analytically, the channel quantization for the case when the nodes A and B use 4-PSK signal set is provided in Section IV C. The channel quantization obtained is shown  to be the same as the one which was obtained in \cite{KoPoTa} using computer search.
\end{itemize}

\section{SINGULAR FADE STATES FOR M-PSK SIGNAL SET}

Throughout the paper the points in the symmetric $M$-PSK signal set are assumed to be of the form $e^{j (2k+1) \pi/M},0 \leq k \leq M-1$ and $M$ is of the form $2^\lambda$, where $\lambda$ is a positive integer.
Let $\Delta\mathcal{S}$ denote the difference constellation of the $M$-PSK signal set $\mathcal{S}$, i.e., $\Delta\mathcal{S}=\lbrace s_i-s'_i \vert  s_i, s'_i \in \mathcal{S}\rbrace$.

For any M-PSK signal set, the set $\Delta\mathcal{S}$ is of the form,

{\footnotesize
\begin{align}
\nonumber
\Delta\mathcal{S}=&\left\lbrace 0\right\rbrace\cup \left\lbrace 2\sin(\pi n /M) e^{j k 2 \pi/M}  \vert{n \; \textrm{odd} }\right\rbrace\\
\nonumber
&\hspace{2 cm}\cup\left\lbrace 2\sin(\pi n /M) e^{j (k 2 \pi/M +  \pi/M)}\vert{n \; \textrm{even} }\right\rbrace,
\end{align}
}where $1 \leq n \leq M/2$ and $0 \leq k \leq M-1$.

\begin{figure}[htbp]
\centering
\includegraphics[totalheight=3.5in,width=6in]{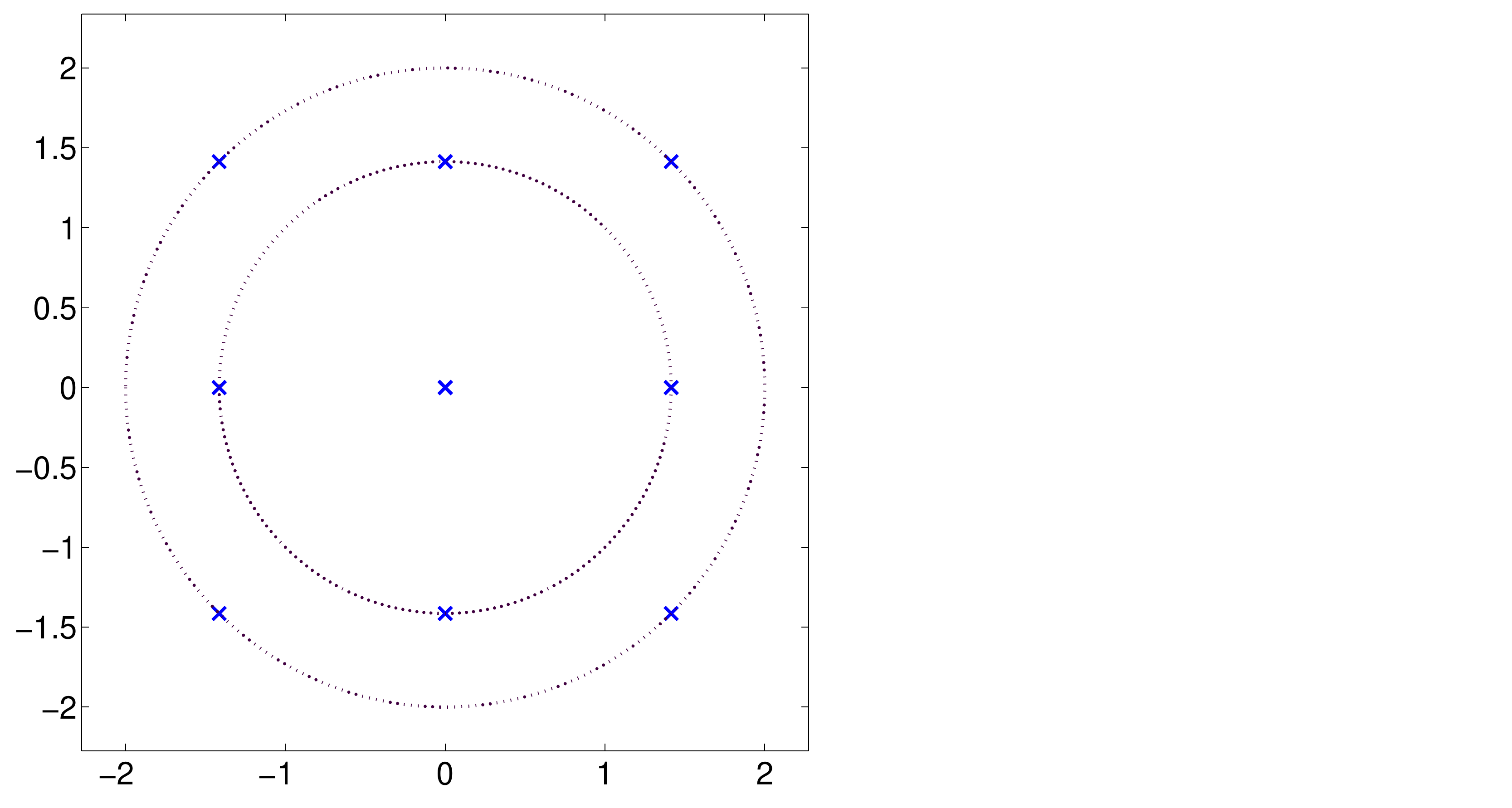}
\caption{Difference constellation for 4-PSK signal set}	
\label{4psk_diff}	
\end{figure}

\begin{figure}[htbp]
\centering
\includegraphics[totalheight=3in,width=5in]{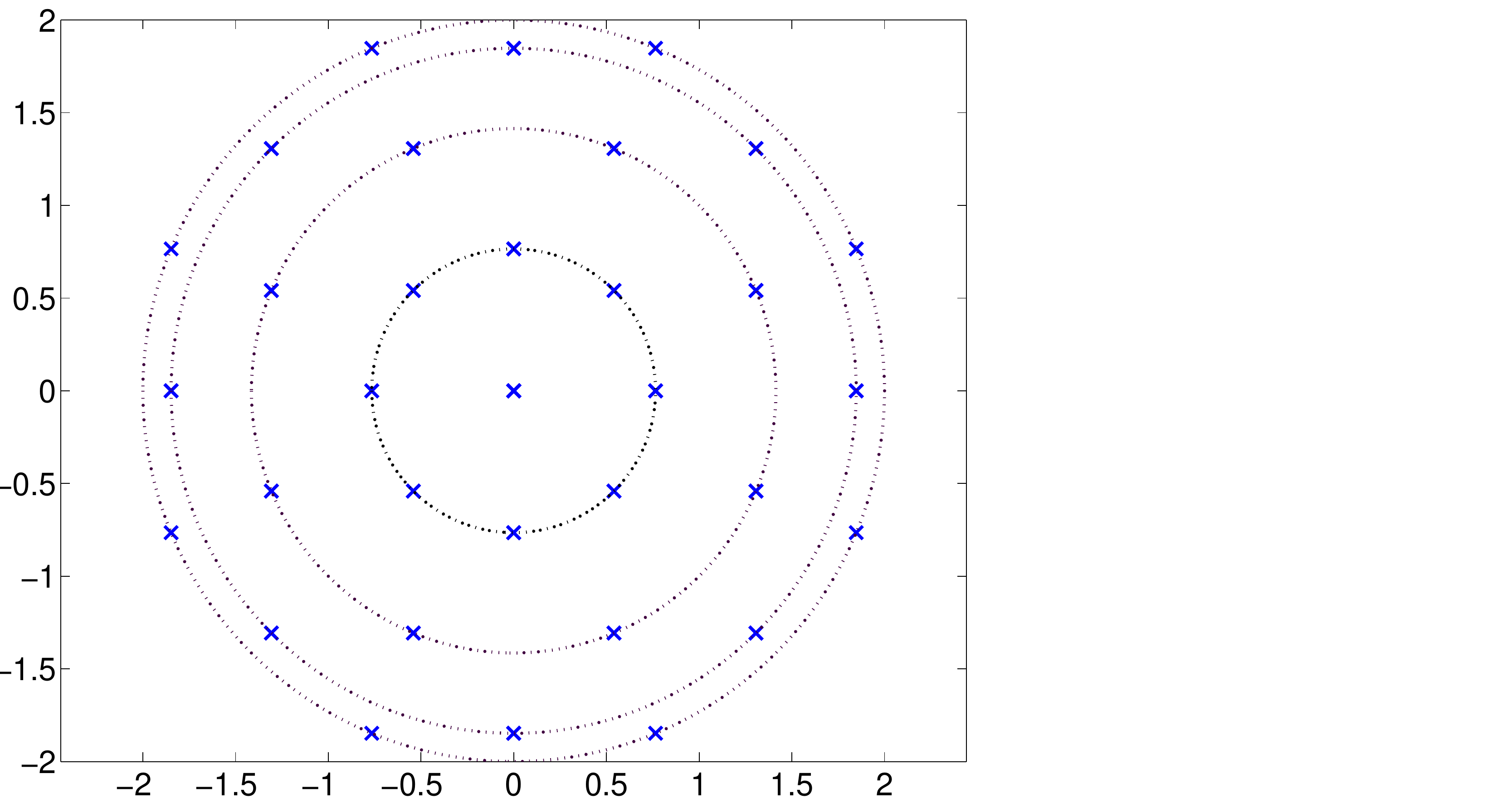}
\caption{Difference constellation for 8-PSK signal set}	
\label{8psk_diff}	
\end{figure}

In other words, the non-zero points in $\Delta\mathcal{S}$ lie on $M/2$ circles of radius $2\sin(\pi n/M), 1 \leq n \leq M/2$ with each circle containing $M$ points. The phase angles of the $M$ points on each circle is $2 k \pi/M$, if $n$ is odd and $2k \pi/M+\pi/M$ if $n$ is even, where $0 \leq k \leq M-1$. For example the difference constellation for 4-PSK and 8-PSK signal sets are shown in Fig. \ref{4psk_diff} and Fig. \ref{8psk_diff} respectively.

Let us define,

\begin{align}
\nonumber
x_{k,n}=\left\lbrace
\begin{array}{ll}
\nonumber
2\sin(\pi n /M) e^{j k 2 \pi/M} {\;\textrm{if}\; n \; \textrm{is} \; \textrm{odd}, }\\
2\sin(\pi n /M) e^{j (k 2 \pi/M +  \pi/M)} {\;\textrm{if}\; n \; \textrm{is} \; \textrm{even}, }
\end{array}
\right.\\
\label{eqn_diff}
\end{align}
where $1 \leq n \leq M/2$ and $0 \leq k \leq M-1$.

From \eqref{eqn_dmin}, it follows that the singular fade states are of the form, 

{\footnotesize
\vspace{-0.2 cm}
\begin{align*}
\gamma_{s} e^{j \theta_{s}}=-x_{k,n}/x_{k',n'}, 
\end{align*}
}for some $x_{k,n},x_{k',n'} \in \Delta\mathcal{S}$.

\begin{lemma}
For integers $k_1$, $k_2$, $l_1$ and $l_2$, where $$1 \leq k_1,k_2,l_1,l_2 \leq \frac{M}{2},k_1 \neq k_2  \textrm{ and }  l_1 \neq l_2,$$

\begin{align}
\nonumber
\dfrac{\sin(k_1 \pi/M)}{\sin(k_2 \pi/M)}=\dfrac{\sin(l_1 \pi/M)}{\sin(l_2 \pi/M)},
\end{align}
if and only if $k_1 = l_1$ and  $k_2 = l_2$.
\begin{proof}
See Appendix A.
\end{proof}
\end{lemma}

\begin{figure}[htbp]
\centering
\includegraphics[totalheight=3in,width=5.5in]{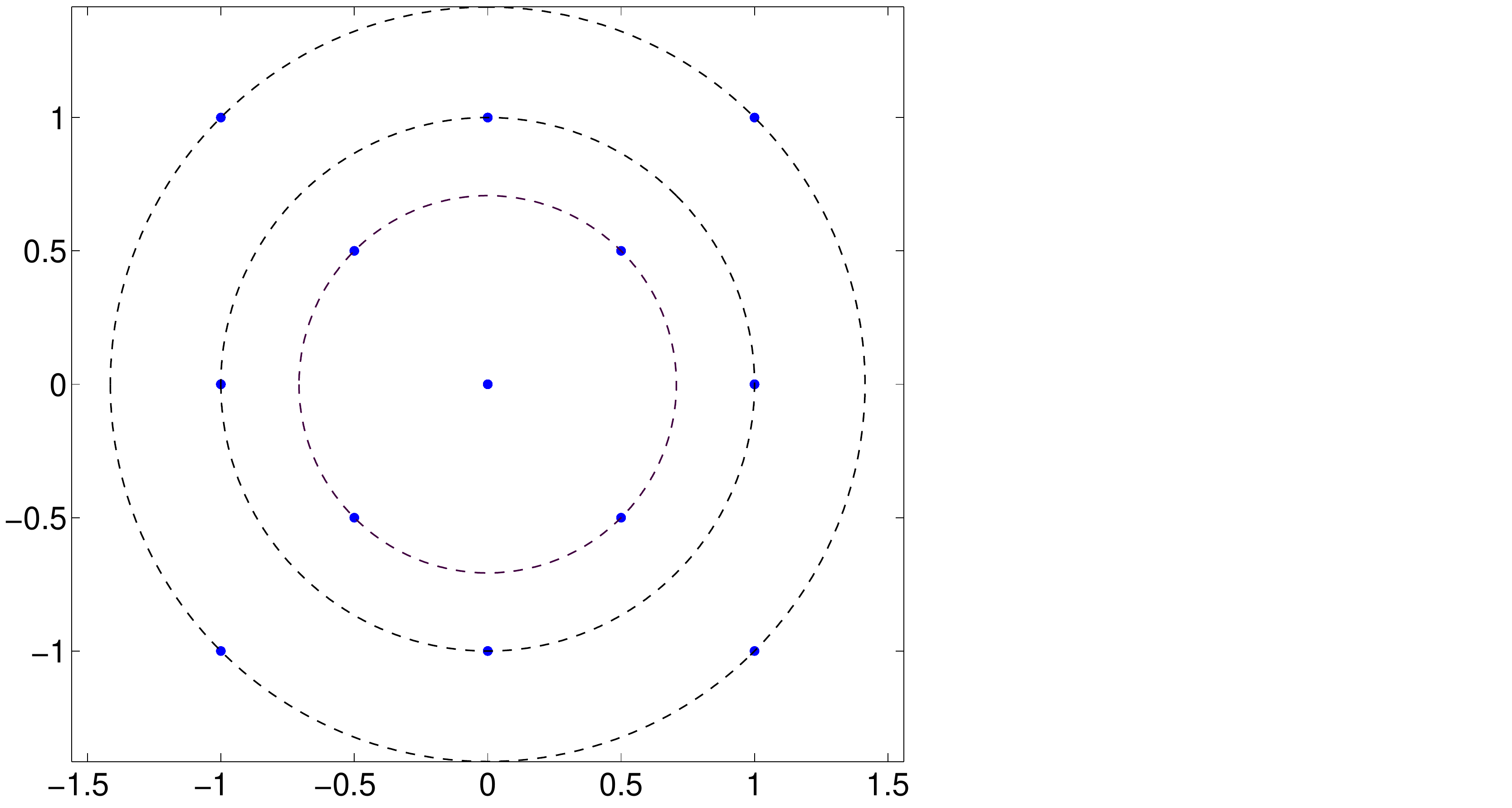}
\caption{Singular fade states for 4-PSK signal set}	
\label{4psk_sing}	
\end{figure}

\begin{figure}[htbp]
\centering
\includegraphics[totalheight=2.5in,width=4.5in]{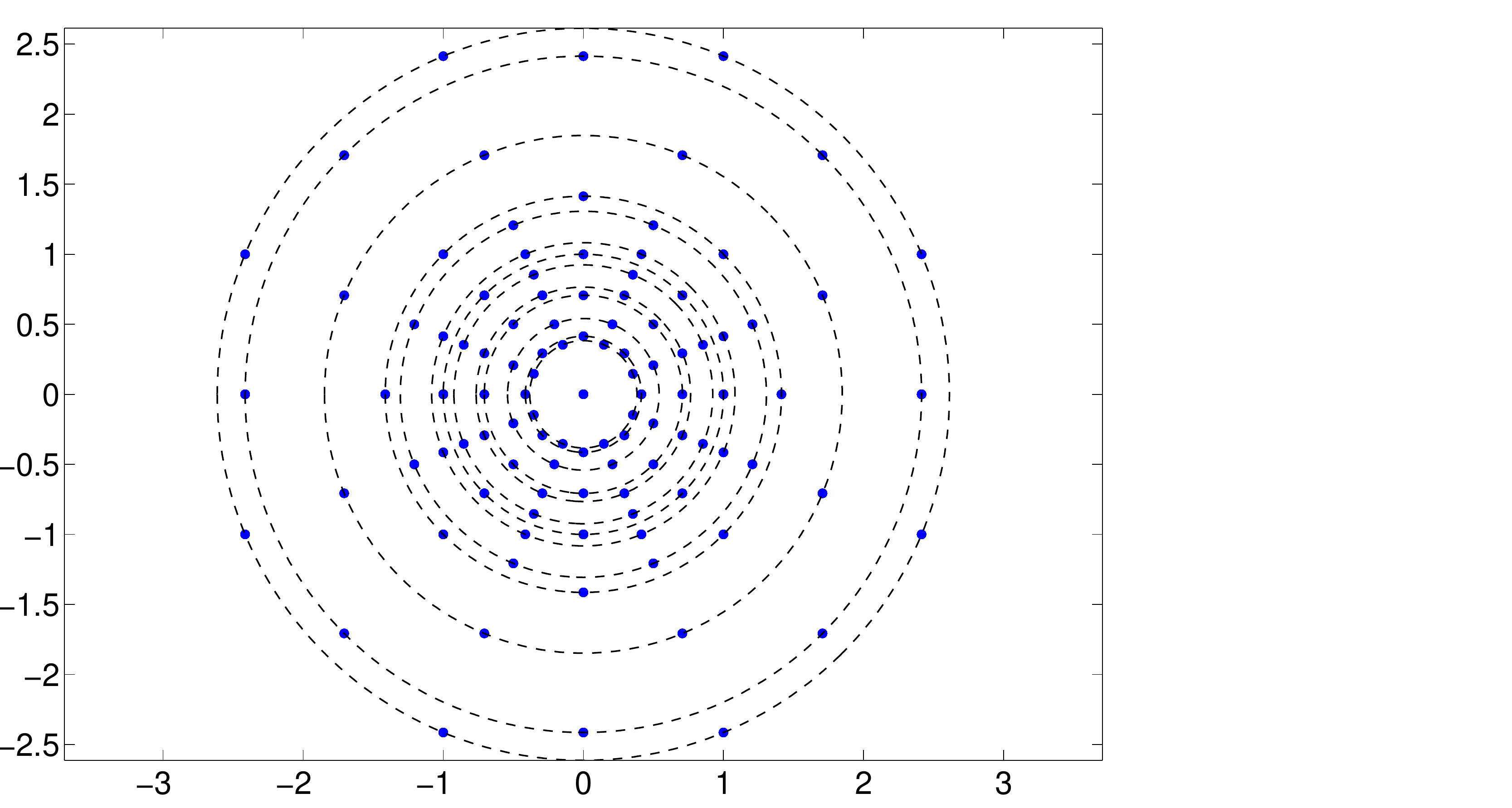}
\caption{Singular fade states for 8-PSK signal set}	
\label{8psk_sing}	
\end{figure}

The following lemma  gives the location of the singular fade states in the complex plane.
\begin{lemma}
\label{lemma_singularity}
The singular fade states other than zero lie on $M^2/4-M/2+1$ circles with $M$ points on each circle, with the radii of the circles given by $\sin(k_1\pi/M)/\sin(k_2 \pi/M)$, where $1 \leq k_1,k_2 \leq M/2$. The phase angles of the $M$ points on each one of the circles are given by $k2\pi/M$, $0 \leq k \leq M-1$, if both $k_1$ and $k_2$ are odd or both are even and $k2\pi/M+\pi/M$, $0 \leq k \leq M-1$, if one among $k_1$ and $k_2$ is odd and the other is even.
\begin{proof}
From \eqref{eqn_diff} it follows that  the amplitude $\gamma_{s}=\sin(\pi n /M)/ \sin(\pi n' /M)$ for some $1 \leq n, n' \leq M/2$ and  the phase $ 2 l\pi/M,0 \leq l \leq M-1$ if  both $n$ and $n'$ are odd or both are even and  $ (2 l+1)\pi/M,0 \leq l \leq M-1$ if one among $n$ and $n'$ is odd and the other is even. To complete the proof of the lemma, it needs to be shown that the number of distinct values possible for $\sin(\pi n /M)/ \sin(\pi n' /M),1 \leq n, n' \leq M/2$ is $M^2/4-M/2+1$. For integers $k_1$, $k_2$, $l_1$ and $l_2$ such that $1 \leq k_1,k_2,l_1,l_2 \leq M/2$, $k_1 \neq k_2$ and  $l_1 \neq l_2$, 

\begin{align}
\nonumber
\dfrac{\sin(k_1 \pi/M)}{\sin(k_2 \pi/M)}=\dfrac{\sin(l_1 \pi/M)}{\sin(l_2 \pi/M)},
\end{align}
if and only if $k_1 = l_1$ and  $k_2 = l_2$. Hence the value of $\sin(\pi n /M)/ \sin(\pi n' /M)$ is distinct whenever $n \neq n'$. Out of the $M^2/4$ values possible for $n,n'$, we subtract out the number of cases for which $n=n'$ which is equal to $M/2$. But then we have to add one to account for all $n=n'$. Hence, we have $M^2/4-M/2+1$ distinct values for $\sin(\pi n /M)/ \sin(\pi n' /M),1 \leq n, n' \leq M/2$.
\end{proof}
\end{lemma}
\begin{example}
For the case when 4-PSK signal set is used during the MA Phase, the singular fade states lie on three circles as shown in Fig. \ref{4psk_sing}.
\end{example}
\begin{example}
For the case when 8-PSK signal set is used during the MA Phase, the singular fade states lie on thirteen circles as shown in Fig. \ref{8psk_sing}.
\end{example}

\section{THE SINGULARITY-FREE REGION}
In this section, it is shown that there exists a set of values for $\gamma e^{j\theta}$ for which any clustering satisfying the exclusive law gives the same minimum cluster distance. Such a region in the complex plane, which contains no singular fade state, is referred to as the singularity-free region. 
\begin{definition}
The set of all values of the fade state (the region in the complex plane) for which any clustering satisfying the exclusive law gives the same minimum cluster distance is defined to be the singularity-free region. The region in the  complex plane other than the singularity-free region is referred to as the singularity region.
\end{definition}

\subsection{Existence of the singularity-free region for $M$-PSK signal set}
The following lemma, which gives an upper bound on the minimum cluster distance, is useful to prove the existence of the singularity-free region for $M$-PSK signal set.
\begin{lemma}
For any clustering $\mathcal{C}^{\gamma,\theta}$ satisfying the exclusive law, with $M$-PSK signal set used at A and B during the MA phase, $d_{min}(\mathcal{C}^{\gamma,\theta})$ is upper bounded as,
\begin{align}
\label{eqn_lemma1}
d_{min}(\mathcal{C}^{\gamma,\theta}) \leq \min\left(2\sin\left(\pi/M\right),2 \gamma \sin\left(\pi/M\right)\right).
\end{align}
\begin{proof}
Since $\mathcal{C}^{\gamma,\theta}$ satisfies the exclusive law, $\mathcal{M}^{\gamma,\theta}(x_A,x_B) \neq \mathcal{M}^{\gamma,\theta}(x_A,x_B')$, where $x_A,x_B,x_B' \in \mathcal{S}$ and $x_B \neq x_B'$. 
From the definition of $d_{min}(\mathcal{C}^{\gamma,\theta})$, we have,

{\footnotesize
\vspace{-0.2cm}
\begin{align}
\nonumber
d_{min}(\mathcal{C}^{\gamma,\theta}) &=\hspace{-0.5cm}\min_{\substack{{{(x_A,x_B) \neq (x'_A,x'_B) \in \mathcal{S}\times\mathcal{S}}} \\{\mathcal{M}^{\gamma,\theta}(x_A,x_B) \neq \mathcal{M}^{\gamma,\theta}(x'_A,x_B')}}}\vert (x_A-x'_A)+\gamma e^{j \theta} (x_B-x'_B)\vert\\
\nonumber
&\leq \min_{{{(x_A,x_B) \neq (x_A,x'_B) \in \mathcal{S}\times\mathcal{S}} }}\vert \gamma e^{j \theta} (x_B-x'_B)\vert\\
\nonumber
&=\gamma\min_{x_B \neq x'_B \in \mathcal{S}}\vert (x_B-x'_B)\vert\\
\label{eqn1}
&=2\gamma \sin\left(\pi/M\right),
\end{align}
}since the minimum distance of the unit energy M-PSK constellation is $2\sin\left(\pi/M\right)$.

Similarly we have $\mathcal{M}^{\gamma,\theta}(x_A,x_B) \neq \mathcal{M}^{\gamma,\theta}(x_A',x_B)$, where $x_A,x_A',x_B \in \mathcal{S}$ and $x_A \neq x_A'$. As a result we have, 
\begin{align}
\label{eqn2}
d_{min}(\mathcal{C}^{\gamma,\theta}) &\leq 2\sin\left(\pi/M\right).
\end{align}
Combining \eqref{eqn1} and \eqref{eqn2}, we get \eqref{eqn_lemma1}.
\end{proof}
\end{lemma}

For any $\gamma e^{j \theta}$ for which  $d_{min}(\gamma e^{j \theta}) = \min\left(2\sin\left(\pi/M\right),2 \gamma \sin\left(\pi/M\right)\right)$, from Lemma 1 and noting the fact that $d_{min}(\mathcal{C}^{\gamma,\theta}) \geq d_{min}(\gamma e^{j \theta})$, we have $d_{min}(\mathcal{C}^{\gamma,\theta}) =\min\left(2\sin\left(\pi/M\right),2 \gamma \sin\left(\pi/M\right)\right)$, for all clusterings $\mathcal{C}^{\gamma,\theta}$ satisfying the exclusive law. In other words, if $\gamma e^{j \theta}$  is such that  $d_{min}(\gamma e^{j \theta}) = \min\left(2\sin\left(\pi/M\right),2 \gamma \sin\left(\pi/M\right)\right)$, all the clusterings satisfying the exclusive law give rise to the same minimum cluster distance. For $\gamma 
\gg 1$, from \eqref{eqn_dmin}, it can be seen that $d_{min}(\gamma e^{j \theta})$ occurs when $x_B=x'_B$ and the value of $d_{min}(\gamma e^{j \theta})=2 \sin(\pi/M)$ and similarly for $\gamma \ll 1$, $d_{min}(\gamma e^{j \theta})$ occurs when $x_A=x'_A$ and the value of $d_{min}(\gamma e^{j \theta})=2 \gamma \sin(\pi/M)$. Hence the region in the complex plane for  which $\gamma \gg 1$ and $\gamma \ll 1$ all clusterings satisfying the exclusive law give rise to the same minimum cluster distance and hence belongs to the singularity-free region. The singularity-free region for $M$-PSK signal set is obtained explicitly in the following subsection. 

\subsection{Singularity-Free Region for M-PSK Signal Set}
\begin{definition}
The outer envelope region formed by a set of circles in the complex plane is defined to be the region exterior to the outer envelope formed by the circles.  
\end{definition}

\begin{figure}[htbp]
\centering
\includegraphics[totalheight=3in,width=2in]{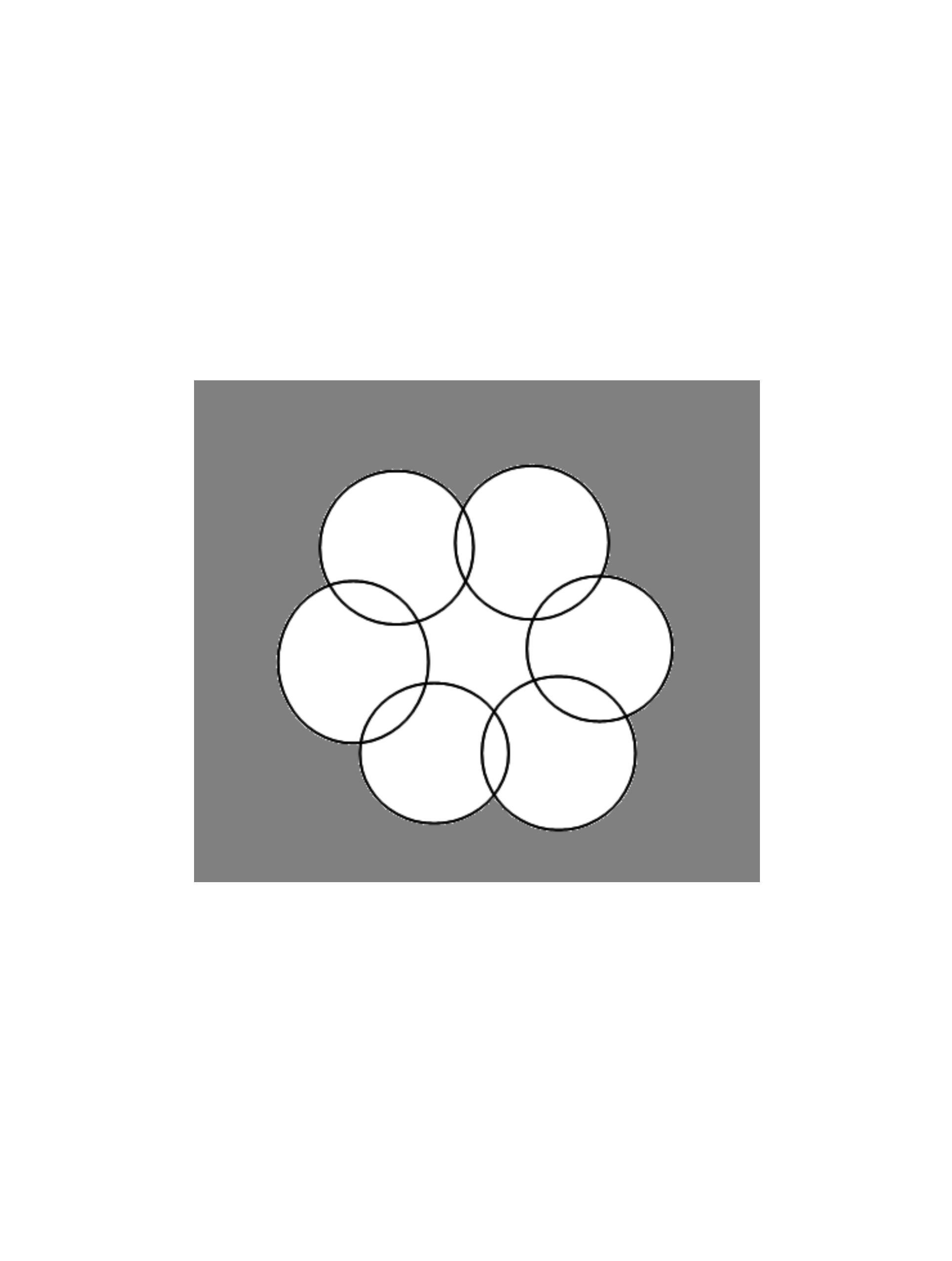}
\vspace{-2 cm}
\caption{Diagram illustrating the outer envelope region formed by a set of circles}	
\label{fig:outer_envelope}	
\end{figure}

For example, the outer envelope region formed by the six circles, is the shaded region shown in Fig. \ref{fig:outer_envelope}.

\vspace{-2 cm}

\begin{definition}
The inner envelope region formed by a point and a set of circles and straight lines in the complex plane is defined to be the interior region formed by the boundaries which are closest to the point.  
\end{definition}

\begin{figure}[htbp]
\centering
\includegraphics[totalheight=2in,width=2in]{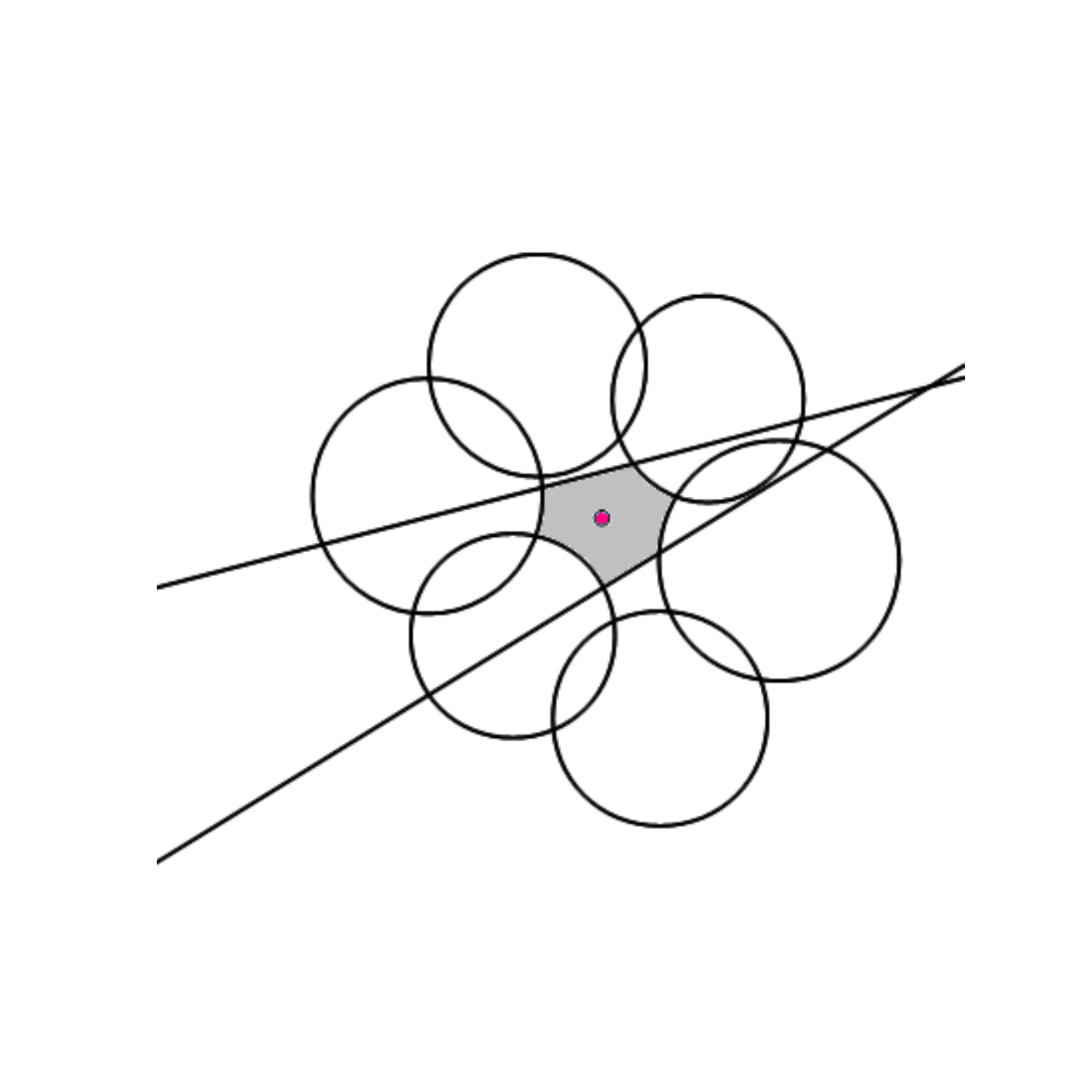}
\vspace{-1 cm}
\caption{Diagram illustrating the inner envelope regions formed by a point and a set of circles and straight lines}	
\label{fig:inner_envelope}	
\end{figure}

For example, the inner envelope region formed by the point, the six circles and the two straight lines, is the shaded region shown in Fig. \ref{fig:inner_envelope}.

The singularity-free region for $M$-PSK signal set is given by the following theorem.

\begin{theorem}
For $M$-PSK signal set, the singularity-free region is the union of the two regions given below. 
\begin{enumerate}

\item 
Region I:

The outer envelope region formed by the $2M$ unit circles with centers at $\cot (\pi/M) e^{jk2\pi/M}$ and $\mathrm{cosec} (\pi/M) e^{j(2k+1)\pi/M}, 0 \leq k \leq M-1$. 
\item
Region II:

\begin{itemize}
\item
For $M>4$, the inner envelope region formed by the origin and  the following $2M$ circles : $M$ circles with centers at $\sec (\pi/M) \tan (\pi/M) e^{j(2k+1)\pi/M}, 0 \leq k \leq M-1$ with radius $\tan^2(\pi/M)$ and $M$ circles with centers at $1/2 \tan (2\pi/M) e^{jk2\pi/M}, 0 \leq k \leq M-1$ with radius $1/2\tan(\pi/M)\tan(2\pi/M)$.
\item
For $M=4$, the inner envelope region formed by the origin and the following four unit circles with centers at $\sqrt{2}e^{j(2k+1)\pi/4},0 \leq k \leq 3$ and the following four straight lines $\gamma e^{j \theta}=\pm 0.5$, $\gamma e^{j \theta}=\pm 0.5j$.
\end{itemize}
\end{enumerate} 
\begin{proof}
See Appendix B.
\end{proof}
\end{theorem}

 Recall from Section II that the singular fade states lie on circles centered at the origin. The centers of the $2M$ circles used to obtain Region I are nothing but the singular fade states which lie on the two outermost circles. From Lemma \ref{sf_int_ext} given in Appendix B, it can be seen that Region II described in Theorem I is the region obtained by complex inversion (the transformation $f(z)=1/z$) of Region I. In other words, if $z \in \mathbb{C}$ is a point in Region I, then $1/z \in \mathbb{C}$ is a point in Region II. 
 
\begin{figure}[htbp]
\centering
\includegraphics[totalheight=3.5in,width=3.5in]{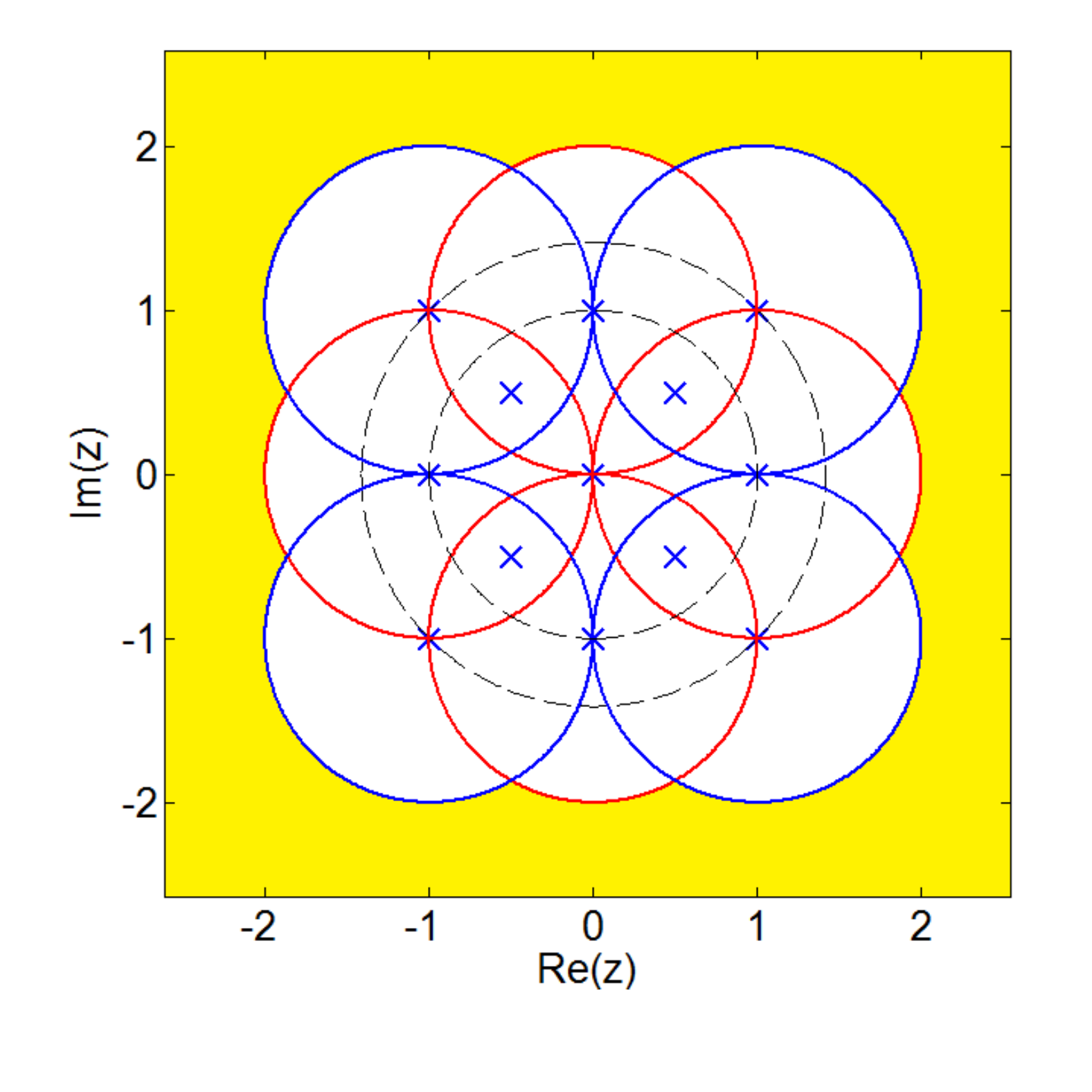}
\caption{Diagram showing Region I described in Theorem 1 for 4-PSK signal set}	
\label{fig:region_outer}	
\end{figure}

\begin{figure}[htbp]
\centering
\includegraphics[totalheight=3.5in,width=3.5in]{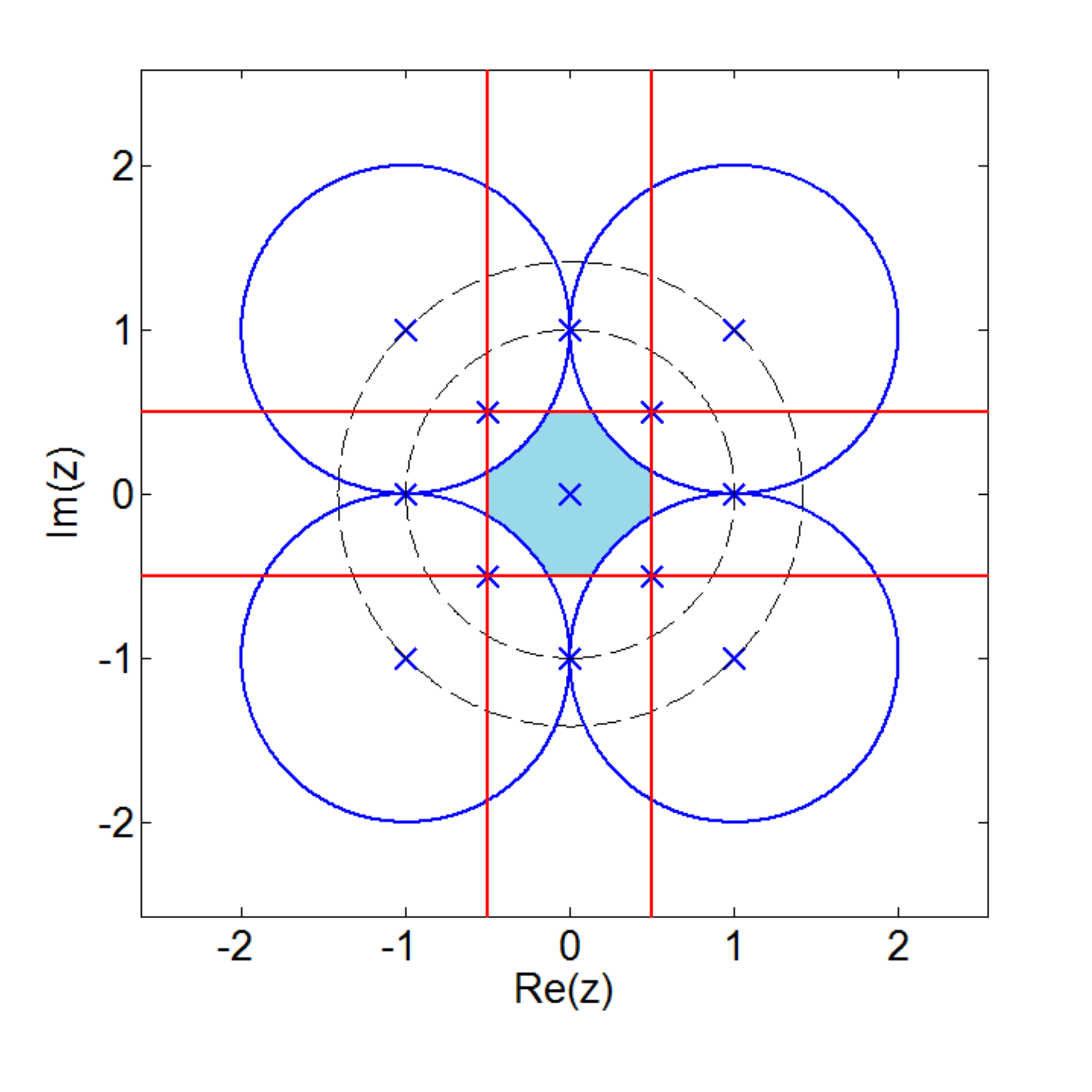}
\caption{Diagram showing Region II described in Theorem 1 for 4-PSK signal set}	
\label{fig:region_inner}	
\end{figure}

\begin{figure}[htbp]
\centering
\includegraphics[totalheight=3.5in,width=3.5in]{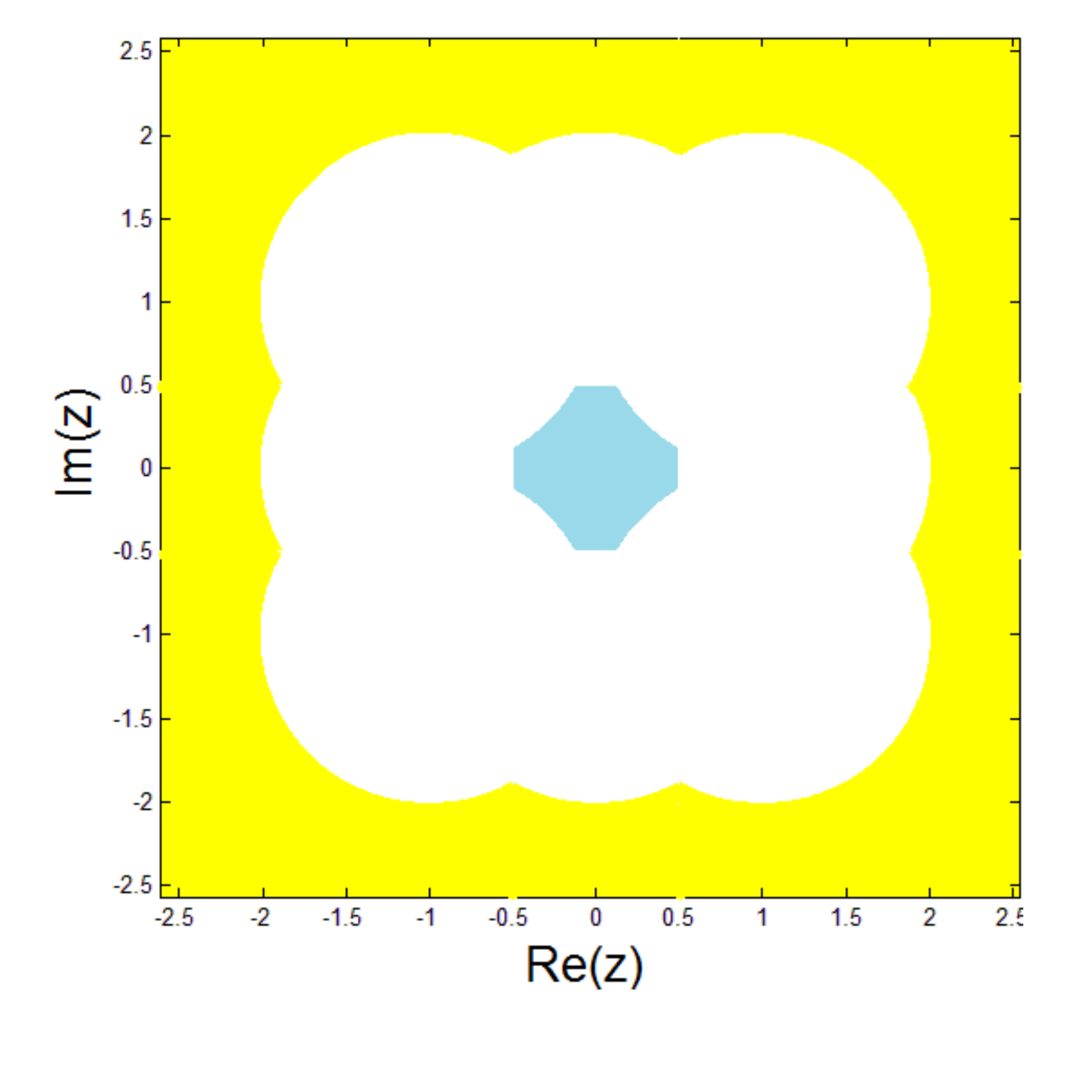}
\caption{Diagram showing the singularity-free region for 4-PSK signal set}	
\label{fig:regions_all}	
\end{figure}

\begin{figure}[htbp]
\centering
\includegraphics[totalheight=3.5in,width=3.5in]{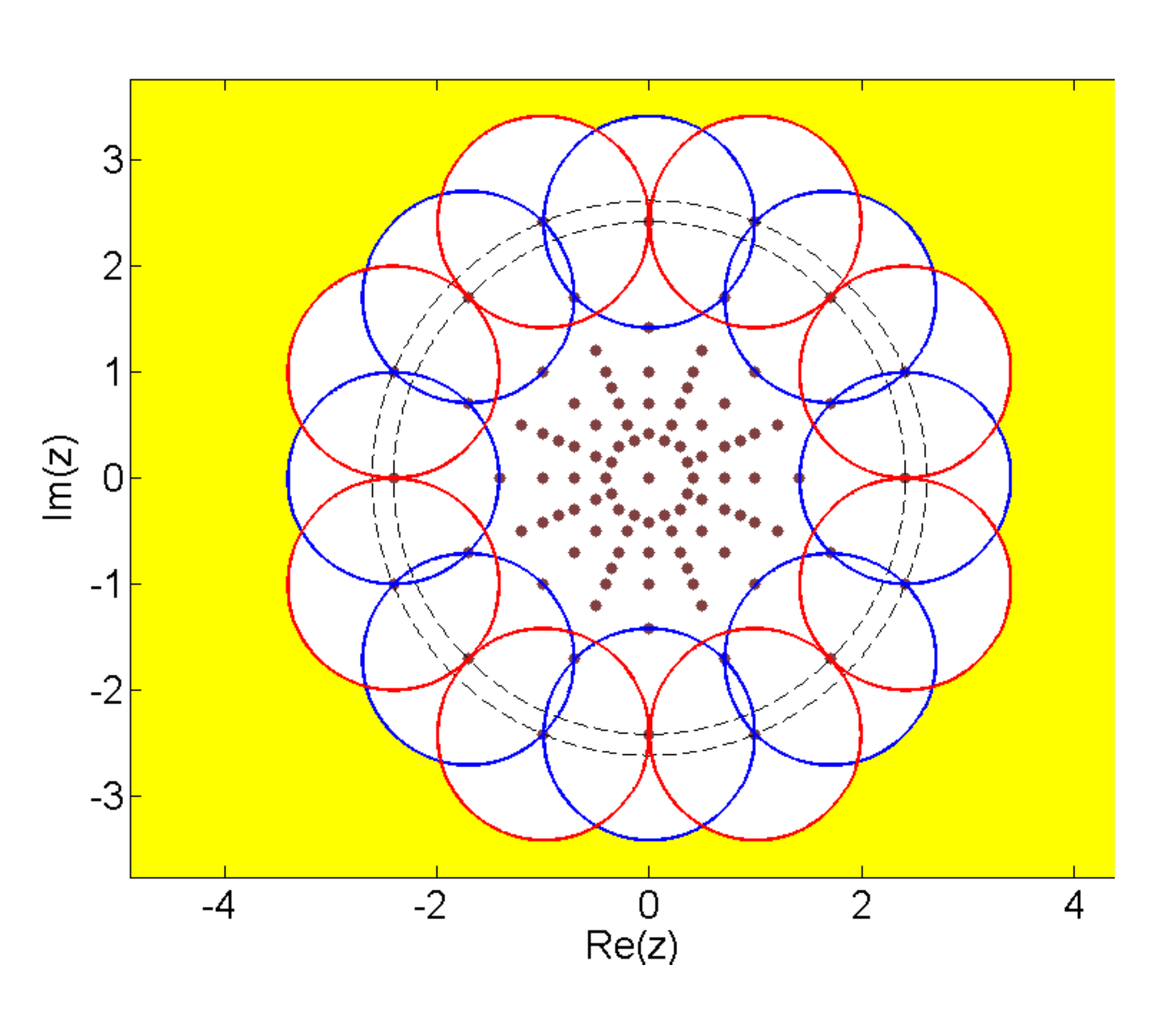}
\caption{Diagram showing Region I described in Theorem 1 for 8-PSK signal set}	
\label{fig:8psk_outer}	
\end{figure}

\begin{figure}[htbp]
\centering
\includegraphics[totalheight=3in,width=3.5in]{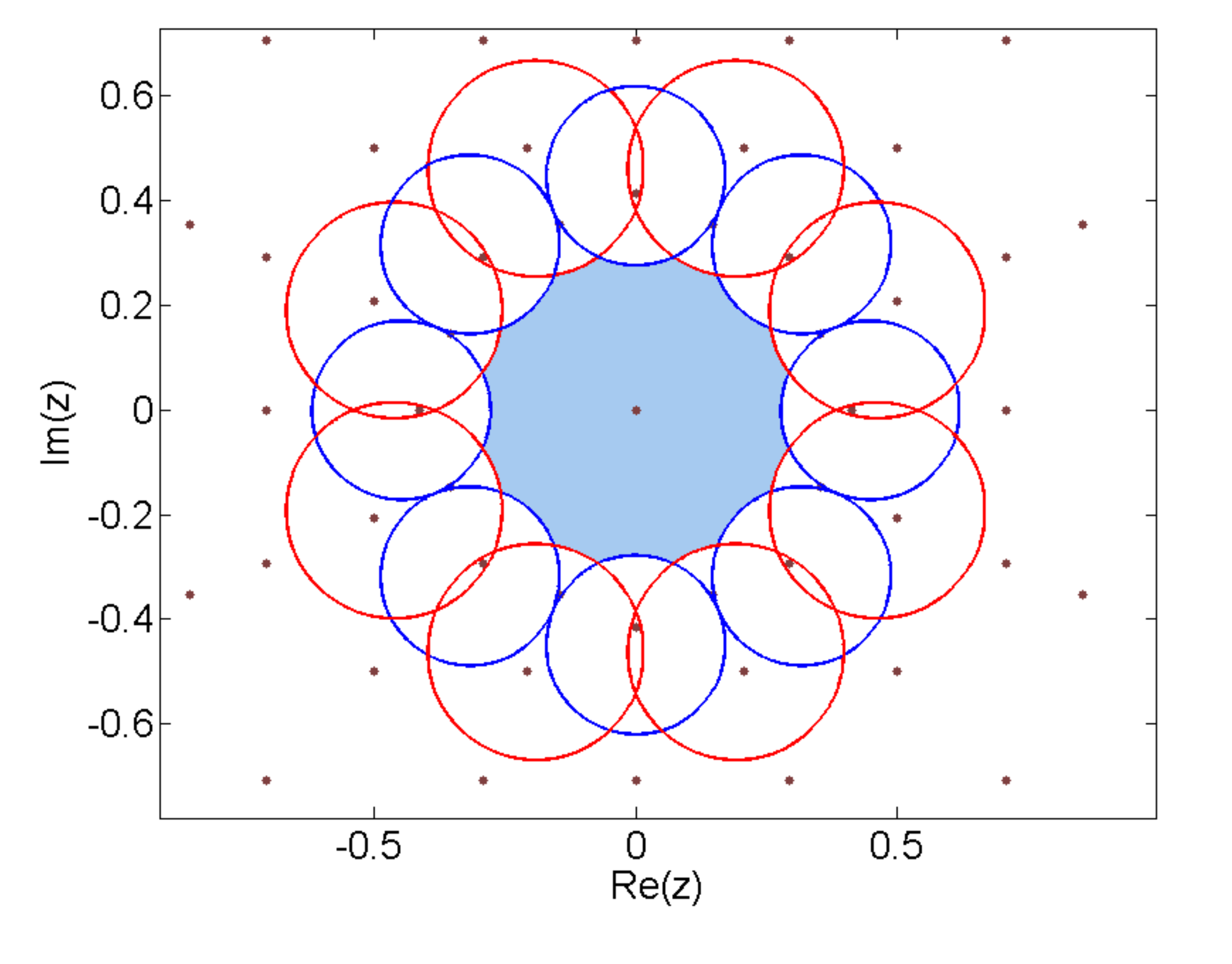}
\caption{Diagram showing Region II described in Theorem 1 for 8-PSK signal set}	
\label{fig:8psk_inner}	
\end{figure}

\begin{figure}[htbp]
\centering
\includegraphics[totalheight=3.5in,width=3.5in]{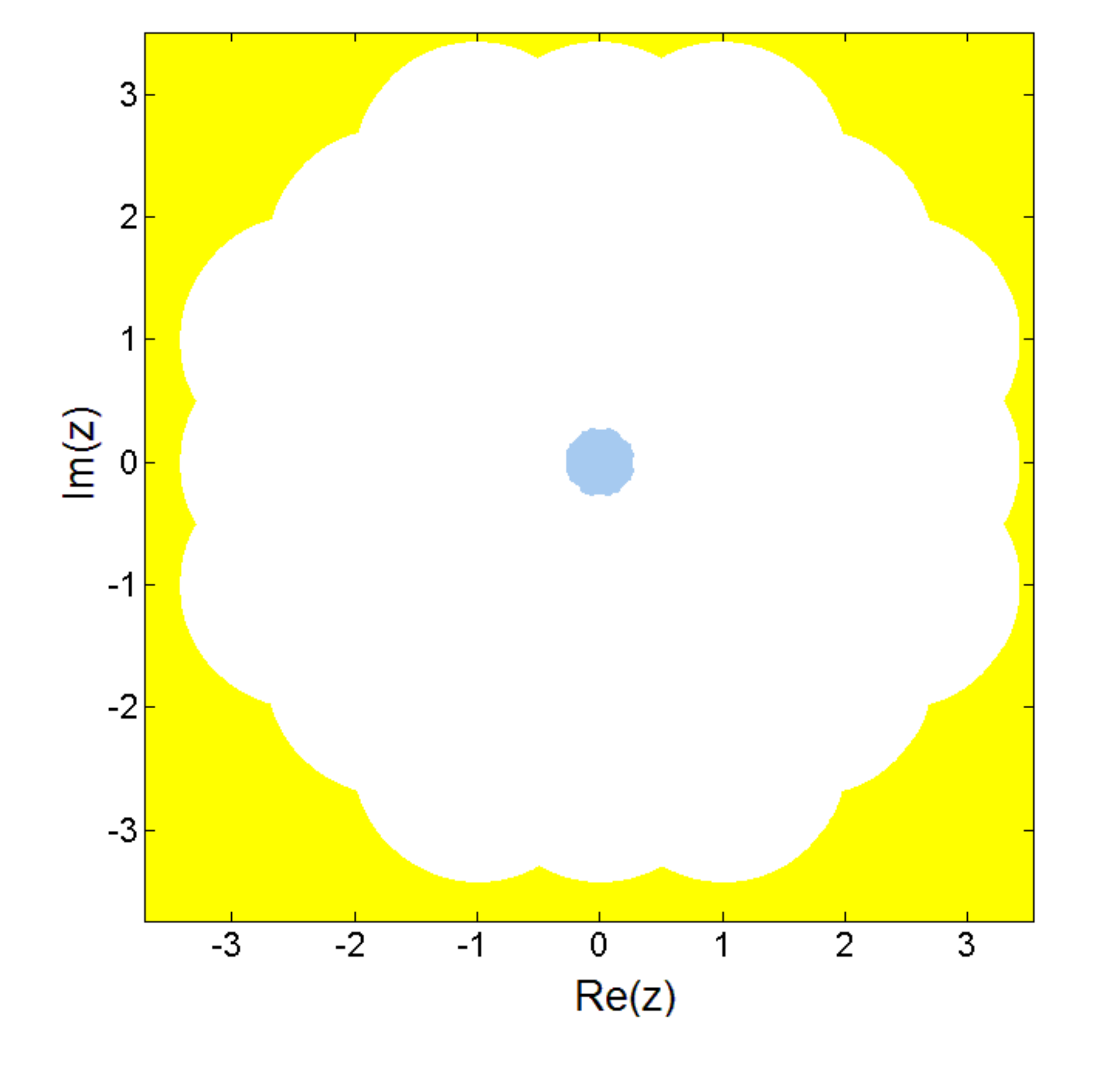}
\caption{Diagram showing the singularity-free region for 8-PSK signal set}	
\label{fig:8psk_all}	
\end{figure}

 \begin{example}
 Consider the case when the nodes A and B use 4-PSK signal set. For this case the 8 unit circles centered at the singular fade states lying on the two outermost circles are shown in Fig. \ref{fig:region_outer}. Region I described in Theorem I is shaded yellow in Fig. \ref{fig:region_outer}. Region II described in Theorem I is shaded blue in Fig. \ref{fig:region_inner}. Note that Region II can also be obtained by the complex inversion of Region I. From Fig. \ref{fig:region_outer} and Fig. \ref{fig:region_inner} we observe some interesting properties of complex inversion \cite{Ne}. 
 \begin{itemize}
 \item
 Circles and straight lines, after complex inversion, become circles and straight lines. The eight circles shown in Fig. \ref{fig:region_outer}, become four circles and four straight lines after complex inversion as shown in Fig. \ref{fig:region_inner}.
 \item
 The circles which touch the origin become straight lines after complex inversion. The four red-colored circles shown in Fig. \ref{fig:region_outer} which touch the origin, after complex inversion become the four straight lines shown in Fig. \ref{fig:region_inner}.
 \item
 The circles which are orthogonal to the unit circle centered at the origin are reflected about the real axis after complex inversion. Hence the four blue-coloreds circles in Fig. \ref{fig:region_outer}, which are orthogonal to the unit circle centered at the origin remain unchanged after complex inversion (since after reflection about the real axis the set of four circles remains the same), as shown in Fig. \ref{fig:region_inner}. 
 \end{itemize}
 
 The union of the yellow and blue regions in Fig. \ref{fig:region_outer} and Fig. \ref{fig:region_inner} respectively, which is the singularity-free region, is shown in Fig. \ref{fig:regions_all}. The white region in Fig. \ref{fig:regions_all} is the singularity region.
 \end{example}
 \begin{example}
 Consider the case when the nodes A and B use 8-PSK signal set. For this case the 16 unit circles centered at the singular fade states lying on the two outermost circles are shown in Fig. \ref{fig:8psk_outer}. Region I described in Theorem I is shaded yellow in Fig. \ref{fig:8psk_outer}. Region II described in Theorem I, which is obtained by the complex inversion of Region I is shaded blue in Fig. \ref{fig:8psk_inner}. The union of the yellow and blue regions in Fig. \ref{fig:8psk_outer} and Fig. \ref{fig:8psk_inner} respectively, which is the singularity-free region, is shown in Fig. \ref{fig:8psk_all}. The white region in Fig. \ref{fig:8psk_all} is the singularity region.
 \end{example}

\section{QUANTIZATION OF THE SINGULARITY REGION AND EXAMPLES}

Recall from Section II that the singular fade states are of the form $-d_k/d_l$, where $d_k,d_l \in \Delta\mathcal{S}$. Throughout this section, we denote a singular fade state by $-d_k/d_l$.

\subsection{A criterion for channel quantization}

 Let $\mathcal{D}(\gamma,\theta)$ denote the set of distances between the points in $\mathcal{S}_R(\gamma,\theta)$, i.e.,

{\footnotesize
\begin{align}
\nonumber
\mathcal{D}(\gamma,\theta)=&\left\lbrace \vert(x_k-x_k')+\gamma e^{j\theta}(x_l-x_l')\vert\right.\\
\nonumber
&\left.\hspace{2.4 cm}:(x_k,x_l) \neq (x_k',x_l') \in \mathcal{S}\times  \mathcal{S} \right \rbrace\\
\nonumber
&=\left\lbrace \vert(d_k)+\gamma e^{j\theta}(d_l)\vert:(d_k,d_l) \neq (0,0) \in \Delta\mathcal{S} \times \Delta\mathcal{S} \right \rbrace
\end{align}
}

The following lemma gives the criterion based on which the singularity region in the complex plane is quantized.
\begin{lemma}
If $\gamma e^{j \theta}$ is such that $\arg \min_{(d_k',d_l')} \mathcal{D}(\gamma,\theta)=(d_k,d_l)$, then the clustering $\mathcal{C}_{\left\lbrace-d_k/d_l\right\rbrace}$ maximizes the minimum cluster distance, among all the clusterings which belong to the set $\mathcal{C}$.
\begin{proof}
Since $\arg \min_{(d_k',d_l')} \mathcal{D}(\gamma,\theta)=(d_k,d_l)$, the minimum distance, $d_{min}(\gamma e^{j \theta})=\vert d_k +  \gamma e^{j \theta}d_l\vert$. Since $\mathcal{C}_{\left\lbrace-d_k/d_l\right\rbrace}$ removes the singular fade state $-d_k/d_l$, $d_{min}^{\mathcal{C}_{\lbrace-d_k/d_l\rbrace}}(\gamma e^{j \theta}) \geq d_{min}(\gamma e^{j \theta})$. The result follows from the fact that for all other clusterings which belong to the set $\mathcal{C}$ which do not remove the singular fade state $-d_k/d_l$, the minimum cluster distance is $d_{min}(\gamma e^{j \theta})$.
\end{proof}
\end{lemma}

Hence, associated with each singular fade state, we have a region in the $\gamma e^{j \theta}$ (complex) plane in which  the clustering which removes that singular fade state maximizes the minimum cluster distance. Let $\mathcal{R}_{\lbrace -d_k/d_l\rbrace}$ denote the region associated with the singular fade state $-d_k/d_l$, i.e.,

{\footnotesize
\begin{align}
\nonumber
\mathcal{R}_{\lbrace -d_k/d_l\rbrace}=&\left\lbrace\gamma e^{j\theta}:\vert d_k+ \gamma e^{j\theta} d_l \vert \leq \vert d_k'+ \gamma e^{j \theta} d_l' \vert ,\right.\\
\label{sing}
&\left.\hspace{0.3 cm} \forall (d_k',d_l') \in \Delta\mathcal{S} \times \Delta\mathcal{S} , (d_k',d_l') \neq (d_k,d_l) \right\rbrace.
\end{align}
} 
\subsection{Quantization of the Singularity Region for M-PSK Signal Set}
\begin{definition}
The pair-wise transition boundary corresponding to a pair of clusterings $c_{\lbrace-d_{k}/d_{l}\rbrace}$ and $c_{\lbrace-d'_{k}/d'_{l}\rbrace}$, denoted by $c({ -d_k/d_l},{ -d_k'/d_l'})$, is the set of values of $\gamma e^{j \theta}$ for which the minimum cluster distances of both the clusterings are equal. In other words, $c({ -d_k/d_l},{ -d_k'/d_l'})$ denotes the curve $\vert d_k+ \gamma e^{j\theta} d_l \vert = \vert d_k'+ \gamma e^{j\theta}d_l' \vert$. 
\end{definition}

The region $\lbrace \gamma e^{j\theta}:\vert d_k+\gamma e^{j \theta} d_l \vert \leq \vert d'_k+\gamma e^{j \theta} d'_l \vert \rbrace$ is the interior region of the curve $c({ -d_k/d_l},{ -d_k'/d_l'})$, if the interior region contains the point $-d_k/d_l$ and it is the exterior region of the curve $c({ -d_k/d_l},{ -d_k'/d_l'})$, if the exterior region contains the point $-d_k/d_l$.

\begin{lemma}
The pair-wise transition boundaries are either circles or straight lines.
\begin{proof}
The curve $\vert d_k+ z d_l \vert = \vert d_k'+ z d_l' \vert$ is obtained by applying the transformation $z'=(d_k+ z d_l)/(d_k'+ z d_l')$ on the curve $\vert z' \vert =1$, which is the unit circle centered at the origin.  The transformation $z'=(d_k+ z d_l)/(d_k'+ z d_l')$ is a linear-fractional transformation \cite{Ne} under which circles become either circles or straight lines. Hence the  pair-wise transition boundaries, which  are the curves obtained by applying a linear-fractional transformation on the unit circle centered at the origin, are either circles or straight lines.
\end{proof}
\end{lemma}

Let $N_S$ denote the number of singular fade states. From \eqref{sing}, it can be seen that in order to obtain the boundaries of the region $\mathcal{R}_{\lbrace -d_k/d_l\rbrace}$, we need to draw $N_S-1$ pairwise transition boundaries $c(-d_k/d_l,-d'_k/d'_l)$, where $-d'_k/d'_l \neq -d_k/d_l \in \mathcal{H}$.

In the following sequel, it is shown that in order to obtain the boundaries of the region $\mathcal{R}_{\lbrace -d_k/d_l\rbrace}$, we need not consider all the $N_S-1$ curves $c({ -d_k/d_l},{ -d_k'/d_l'})$.

 In order to obtain all the regions $\mathcal{R}_{\lbrace -d_k/d_l\rbrace}, -d_k/d_l \in \mathcal{H}$, it is enough to consider the regions corresponding to those singular fade states which lie on the lines $\theta = 0$ and $\theta =\pi/M$. The regions corresponding to the other singular fade states can be obtained by symmetry. In other words, without loss of generality we can assume that the singular fade state $-d_k/d_l$ lies on the line $\theta = 0$ or $\theta=\pi/M$. 
 
 \begin{lemma}
 \label{reg_comp_inv}
 The region $\mathcal{R}_{\lbrace -d_l/d_k\rbrace}, -d_l/d_k \in \mathcal{H}$, is the region obtained by the complex inversion of the region $\mathcal{R}_{\lbrace -d_k/d_l\rbrace}, -d_k/d_l \in \mathcal{H}$.
 \begin{proof}
 By definition, any $\gamma e^{j \theta} \in \mathcal{R}_{\lbrace -d_l/d_k\rbrace}$ should satisfy,
 $\vert d_l+\gamma e^{j\theta}d_k \vert \leq \vert d'_l+\gamma e^{j\theta}d'_k \vert, \forall (d_l',d_k') \neq (d_l,d_k) \in \Delta\mathcal{S} \times \Delta\mathcal{S}$. Equivalently, we have,

{\footnotesize
\begin{align} 
\nonumber
\mathcal{R}_{\lbrace -d_l/d_k\rbrace}=\left\lbrace\gamma e^{j \theta} :
  \left\vert \dfrac{1}{\gamma} e^{-j\theta}d_l+d_k \right\vert \leq \left\vert \dfrac{1}{\gamma} e^{-j\theta}d'_l+d'_k \right\vert\right.\\
  \label{region_inv}
   \left.\forall (d_l',d_k') \neq (d_l,d_k) \in \Delta\mathcal{S} \times \Delta\mathcal{S}\right\rbrace.
  \end{align}
}

Comparing \eqref{sing} and \eqref{region_inv}, we get the result.
 
 \end{proof}
 \end{lemma}
 
 From Lemma \ref{reg_comp_inv}, it follows that it is enough if we obtain the region $\mathcal{R}_{\lbrace -d_k/d_l\rbrace}$ corresponding to those singular fade states which lie outside the unit circle centered at the origin. The regions corresponding to those singular fade states which lie inside the unit circle centered at the origin can be obtained by complex inversion. In the rest of this subsection, it is assumed that the region $\mathcal{R}_{\lbrace -d_k/d_l\rbrace}$ to be obtained is associated with a singular fade state $-d_k/d_l$ which lies outside the unit circle.
 
 For example, for the case when 4-PSK signal set is used at A and B, it is enough if we obtain the regions corresponding to those singular fade states which lie on the circle with radius $\sqrt{2}$ (see Fig. \ref{4psk_sing}). The regions corresponding to those singular fade states which lie on the circle with radius $1/\sqrt{2}$ can be obtained by complex inversion of the regions corresponding to those singular fade states which lie on the circle with radius $\sqrt{2}$. We need not explicitly draw the boundaries of the regions corresponding to those singular fade states which lie on the unit circle. The boundaries of the regions corresponding to those singular fade states which lie on the unit circle are automatically formed once the regions corresponding to all other singular fade states are drawn.
 
\begin{figure}[htbp]
\centering
\includegraphics[totalheight=3in,width=3.75in]{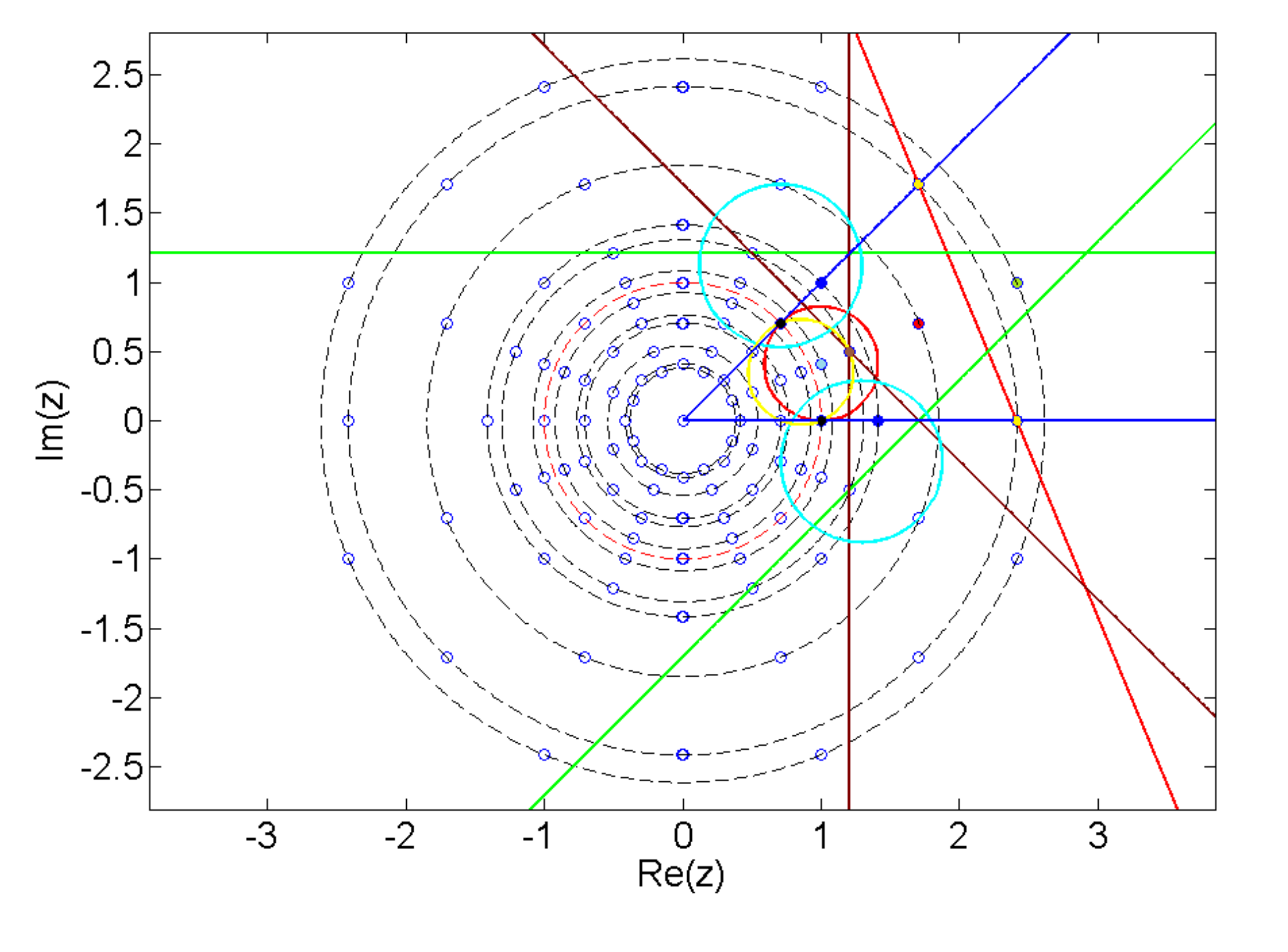}
\caption{Diagram showing the pair-wise transition boundaries corresponding to the singular fade state $1.8478e^{j \pi/8}$ for 8-PSK signal set}	
\label{8psk_sameline_region}	
\end{figure}

\begin{lemma}
\label{lemma_wedge}
The region $R_{\lbrace -d_k/d_l\rbrace}$, where the singular fade state $-d_k/d_l$ lies on the line $\theta=a, a\in \lbrace0,\pi/M\rbrace$, lies inside the wedge formed by the lines $\theta = a-\pi/M$ and $\theta=a+\pi/M$.
\begin{proof}
Consider the two singular fade states $-d_{k_1}/d_{l_1}=-d_k/d_l e^{j 2 \pi/M}$ and $-d_{k_2}/d_{l_2}=-d_k/d_l e^{-j 2 \pi/M}$. The curve $\vert d_k+\gamma e^{j \theta}d_l\vert=\vert d_{k_1}+\gamma e^{j \theta}d_{l_1}\vert$ is the straight line $\theta=\pi/M$ and for $\theta > \pi/M$, $\vert d_k+\gamma e^{j \theta}d_l>\vert d_{k_1}+\gamma e^{j \theta}d_{l_1}\vert$. Similarly, the curve $\vert d_k+\gamma e^{j \theta}d_l\vert=\vert d_{k_2}+\gamma e^{j \theta}d_{l_2}\vert$ is the straight line $\theta=-\pi/M$ and for $\theta < -\pi/M$, $\vert d_k+\gamma e^{j \theta}d_l>d_{k_2}+\gamma e^{j \theta}d_{l_2}\vert$. Hence, from the  mentioned facts and from \eqref{sing}, it follows that $R_{\lbrace -d_k/d_l\rbrace}$ should lie inside the wedge formed by the lines $\theta = a-\pi/M$ and $\theta=a+\pi/M$.
\end{proof}
\end{lemma}
For example, for the case when 8-PSK signal set is used at A and B, the region $\mathcal{R}_{\lbrace1.8478e^{j \pi/8}\rbrace}$ (the singular fade state $1.8478e^{j \pi/8}$ is shown by a red dot in Fig. \ref{8psk_sameline_region}), lies inside the wedge formed by the straight lines $\theta=0$ and $\theta=\pi/4$ (the straight lines $\theta=0$ and $\theta=\pi/4$ are indicated by blue color in Fig. \ref{8psk_sameline_region}).

\begin{lemma}
In order to obtain the boundaries for $\mathcal{R}_{\lbrace -d_k/d_l\rbrace}$, where the singular fade state $-d_k/d_l$ lies on the line $\theta=a, a \in \lbrace0,\pi/M\rbrace$, it is enough to consider the curves $c(-d_k/d_l,-d'_k/d'_l)$, for which the singular fade states $-d'_k/d'_l$ lie on the lines $\theta =a-\pi/M$,  $\theta =a$ and $\theta = a+\pi/M$.
\begin{proof}
The proof is given for the case when $a=0$. The proof for the case when $a=\pi/M$ is similar and is omitted. Consider the singular fade states which lie on the circle with radius $\gamma'$ and have phase angles $2m\pi/M, 0 \leq m \leq M-1$. Let $-d_{k_m'}/d_{l_m'}$ denote the singular fade state whose phase angle is $2m'\pi/M$. It can be verified that for $\theta \leq m' \pi/M$, $\vert d_{k_0}+\gamma e^{j\theta} d_{l_0} \vert \leq  \vert d_{k_m'}+\gamma e^{j\theta} d_{l_m'}\vert$, for $1 \leq m' \leq M/2-1$. Hence inside the wedge formed by the straight lines $\theta=\pi/M$ and $\theta=-\pi/M$, $\vert d_{k_0}+\gamma e^{j\theta} d_{l_0} \vert \leq  \vert d_{k_m'}+\gamma e^{j\theta} d_{l_m'}\vert$, for $1 \leq m' \leq M/2-1$. Similarly it can be shown that inside the wedge formed by the straight lines $\theta=\pi/M$ and $\theta=-\pi/M$, $\vert d_{k_1}+\gamma e^{j\theta} d_{l_1} \vert \leq  \vert d_{k_m'}+\gamma e^{j\theta} d_{l_m'}\vert$, for $M/2 \leq m' \leq M-1$. Hence by Lemma \ref{lemma_wedge}, it is enough if we consider only the singular fade state  $-d_{k_0}/d_{l_0}$, among all the singular fade states which lie on the circle $\gamma'$. Following a similar procedure, it can be shown that among all the singular fade states which lie on a circle with radius $\gamma''$ and have phase angles $(2m+1)\pi/M, 0 \leq m \leq M-1$, it is enough to consider the singular fade states with phase angles $-\pi/M$ and $\pi/M$.
\end{proof}
\end{lemma}

For example, for the case when 8-PSK signal set is used at A and B, to obtain the region $\mathcal{R}_{\lbrace1.8478e^{j \pi/8}\rbrace}$, it is enough to consider the pair-wise transition boundaries  $c(1.8478e^{j \pi/8},-d_k/d_l)$, for which the singular fade states $-d_k/d_l$ lie on the lines $\theta=0$, $\theta=\pi/8$ and $\theta=2\pi/8$. 
\begin{lemma}
\label{lemma:same_line}
Consider the region $\mathcal{R}_{\lbrace -d_k/d_l\rbrace}$, where  $-d_k/d_l$ lies on the line $\theta =a, a \in \lbrace{0,\pi/M}\rbrace$ . In order to obtain the boundaries for $\mathcal{R}_{\lbrace -d_k/d_l\rbrace}$, among all the curves $c( -d_k/d_l,-d'_k/d'_l)$, where $-d'_k/d'_l$ lies on the $\theta=a$ line, it is enough to consider the curve $c( -d_k/d_l,-d_l/d_k)$ (which is the unit circle centered at the origin) and those curves for which $\vert d'_k/d'_l \vert\geq 1$. If there is no singular fade state $-d'_{k}/d'_{l}$ on the line $\theta=a$ which satisfies $\vert d'_{k}/d'_{l} \vert >\vert d_k/d_l\vert$, in addition consider the unit circle centered at $-d_k/d_l$. 
\begin{proof}
To prove the lemma, we first prove that $\mathcal{R}_{\lbrace-d_k/d_l\rbrace}$ lies outside the unit circle $\gamma=1$, for $\vert d_k/d_l \vert >1$. It can be verified that the curve $d_k+\gamma e^{j \theta}d_l$ is the unit circle $\gamma =1 $ and $\vert d_k +\gamma e^{j \theta}d_l \vert>\vert d_l +\gamma e^{j \theta}d_k \vert$ for $\gamma < 1$. Hence from the definition of $\mathcal{R}_{\lbrace-d_k/d_l\rbrace}$ given in \eqref{sing}, it follows that $\gamma <1$ does not belong to $\mathcal{R}_{\lbrace-d_k/d_l\rbrace}$.

Consider a singular fade state $-d_{k_1}/d_{l_1} \neq -d_k/d_l$ with $\vert d_{k_1}/d_{l_1}\vert >1$. Since $\vert d_{k_1} +\gamma e^{j \theta}d_{l_1} \vert \leq \vert d_{l_1} +\gamma e^{j \theta}d_{k_1} \vert$ for $\gamma \geq 1$ and $\mathcal{R}_{\lbrace-d_k/d_l\rbrace}$ lies outside the unit circle $\gamma =1$, we need not consider the singular fade state $-d_{l_1}/d_{k_1}$. From the above argument, it follows that if we consider only the pair-wise transition curves formed by the singular fade state  $-d_{k}/d_{l}$ with those singular fade states which lie on or outside the circle $\gamma=1$. In addition, we need to consider the curve $\vert d_{k} +\gamma e^{j \theta}d_{l} \vert= \vert d_{l} +\gamma e^{j \theta}d_{k} \vert$, which is the unit circle centered at the origin.

If there is no singular fade state $-d'_{k}/d'_{l}$ on the line $\theta=a$ which satisfies $\vert d'_{k}/d'_{l} \vert >\vert d_k/d_l\vert$, from Theorem 1, in order to construct the boundary $\mathcal{R}_{\lbrace-d_k/d_l\rbrace}$ forms with the singularity-free region, we need to consider the unit circle centered at the singular fade state $-d_k/d_l$.
\end{proof}
\end{lemma}
For example, consider the case when 8-PSK signal set is used at A and B. To obtain the region $\mathcal{R}_{\lbrace1.8478e^{j \pi/8}\rbrace}$, from Lemma \ref{lemma:same_line}, it is enough to consider the curves $c(1.8478e^{j \pi/8},2.6131e^{j \pi/8})$, $c(1.8478e^{j \pi/8},1.3066e^{j \pi/8})$, $c(1.8478e^{j \pi/8},1.0824e^{j\pi/8})$ and the unit circle centered at the origin (the singular fade states $2.6131e^{j \pi/8}$, $1.3066e^{j \pi/8}$  and $1.0824e^{j\pi/8}$ are shown respectively by green, brown and light blue dots in Fig. \ref{8psk_sameline_region}). The curves $c(1.8478e^{j \pi/8},2.6131e^{j \pi/8})$, $c(1.8478e^{j \pi/8},1.3066e^{j \pi/8})$ and $c(1.8478e^{j \pi/8},1.0824e^{j\pi/8})$ are respectively the red straight line, the red circle and the yellow circle shown in Fig. \ref{8psk_sameline_region}. The unit circle centered at the origin is the dotted red circle shown in Fig. \ref{8psk_sameline_region}.
\begin{lemma}
\label{lemma:same_pi_by_m}
In order to obtain the region $\mathcal{R}(-d_k/d_l)$, where  $-d_k/d_l$ lies on the line $\theta =a,a \in\lbrace 0, \pi/M \rbrace$ , among all the curves $c( -d_k/d_l,-d'_k/d'_l)$, where $-d'_k/d'_l$ lies on the line $\theta=a', a' \in \lbrace a-\pi/M,a+\pi/M \rbrace$ , it is enough to consider the curves for which $\vert d_{k'}/d'_{l} \vert \geq 1$.
\begin{proof}
The  proof is similar to the proof of Lemma \ref{lemma:same_line} and is omitted. 
\end{proof}
\end{lemma}

 \begin{figure}[htbp]
\centering
\includegraphics[totalheight=3in,width=3.75in]{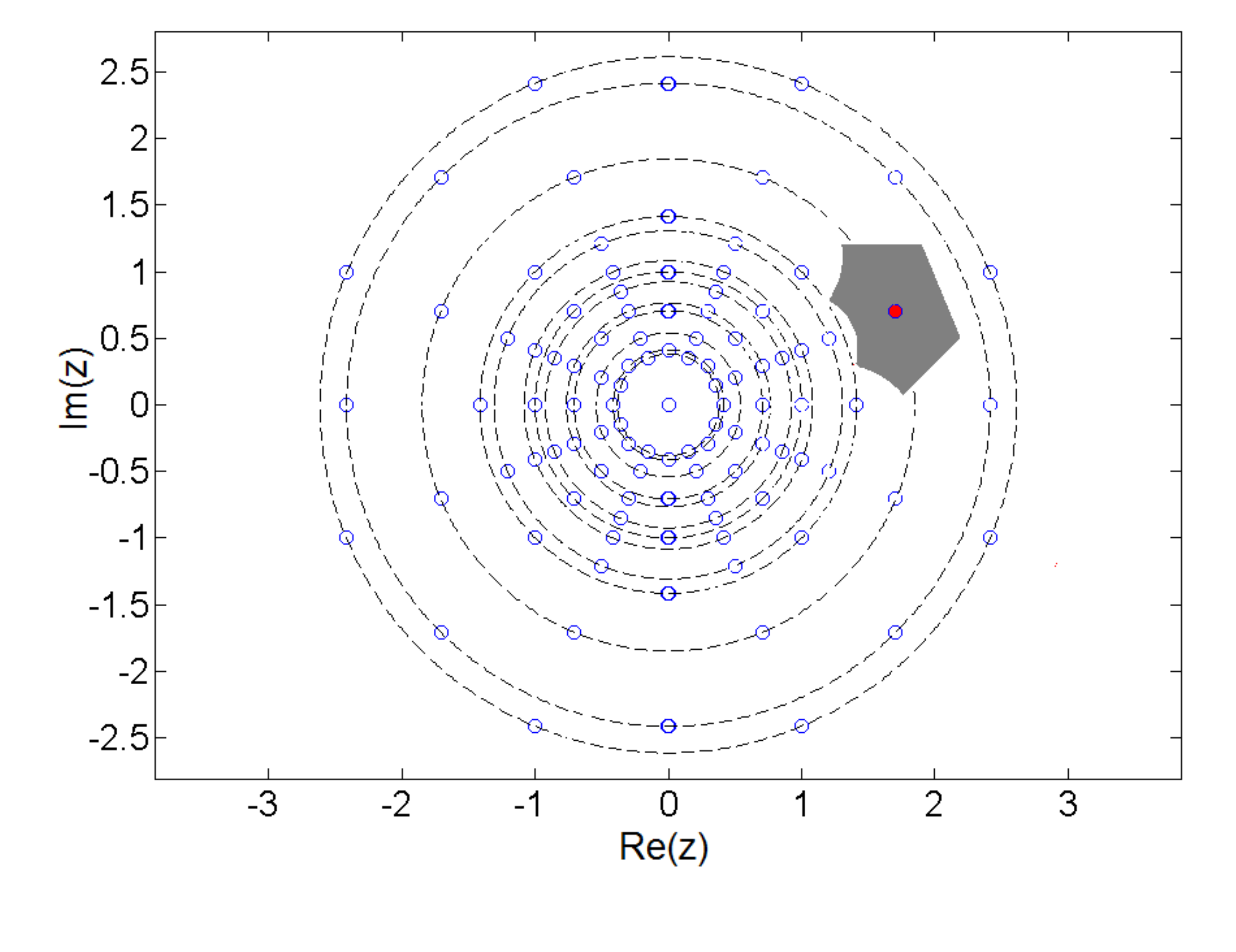}
\caption{Diagram showing the region $\mathcal{R}_{\lbrace1.8478e^{j \pi/8}\rbrace}$ for 8-PSK signal set}	
\label{8psk_region}	
\end{figure}

For example, for the case when 8-PSK signal set is used at A and B, to obtain the region $\mathcal{R}_{\lbrace1.8478e^{j \pi/8}\rbrace}$, it is enough to consider the following pairs of curves $c(1.8478e^{j \pi/8},1)$ and $c(1.8478e^{j \pi/8},e^{j2\pi/8})$ (indicated by brown straight lines in Fig. \ref{8psk_sameline_region}),  $c(1.8478 e^{j \pi/8},1.4142)$ and $c(1.8478 e^{j \pi/8},1.4142e^{j2\pi/8})$ (indicated by blue circles in Fig. \ref{8psk_sameline_region}) and $c(1.8478e^{j \pi/8},2.4142)$ and $c(1.8478e^{j \pi/8},2.4142e^{j2\pi/8})$ (indicated by green straight lines in Fig. \ref{8psk_sameline_region}). In Fig. \ref{8psk_sameline_region}, the singular fade states $1$ and  $e^{j2\pi/8}$ are shown by black dots, the the singular fade states $1.4142$ and $1.4142e^{j2\pi/8}$ are shown by blue dots and the the singular fade states $2.4142$ and $2.4142e^{j2\pi/8}$ are shown by yellow dots.
\begin{theorem}
The region $\mathcal{R}_{\lbrace-d_k/d_l\rbrace}$ , where the singular fade state $-d_k/d_l$ lies on the line $\theta=a$, is the inner envelope region formed by the singular fade state $-d_k/d_l$ and the curves given in Lemma \ref{lemma:same_line}, Lemma \ref{lemma:same_pi_by_m}, and the straight lines $\theta=a-\pi/M$ and $\theta=a+\pi/M$.
\begin{proof}
The proof follows from Lemmas \ref{lemma_wedge}-\ref{lemma:same_pi_by_m}.
\end{proof}
\end{theorem}

For example, the region $\mathcal{R}_{\lbrace1.8478e^{j \pi/8}\rbrace}$ is obtained by finding the inner envelope region formed by the singular fade state $1.8478e^{j \pi/8}$ (shown by a red dot in Fig. \ref{8psk_sameline_region})  and the curves shown in Fig. \ref{8psk_sameline_region}. The obtained region $\mathcal{R}_{\lbrace1.8478e^{j \pi/8}\rbrace}$ is the shaded region shown in Fig. \ref{8psk_region}.

\subsection{Channel Quantization for 4-PSK Signal Set}

\begin{figure}[htbp]
\centering
\includegraphics[totalheight=2.5in,width=5in]{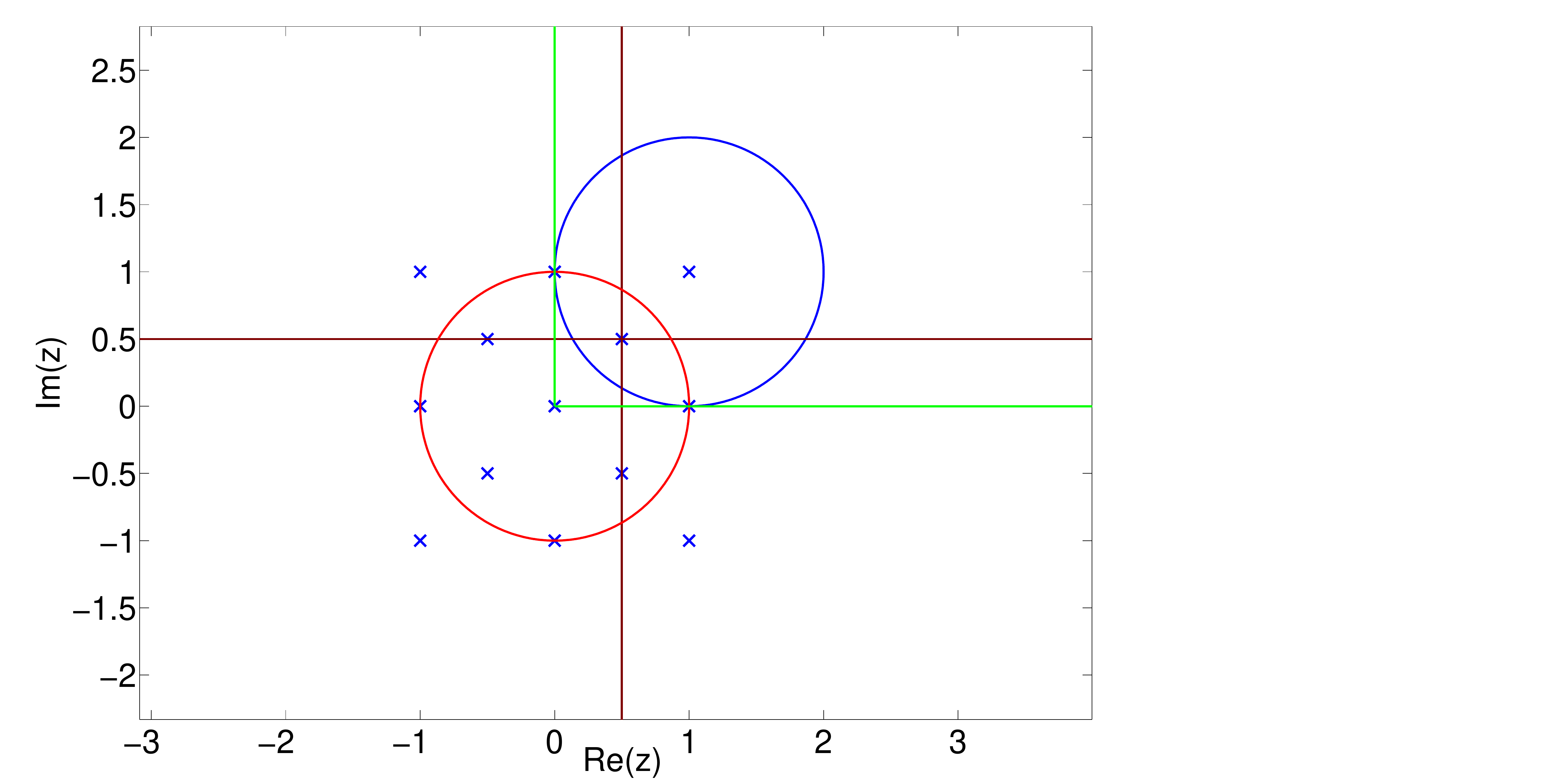}
\caption{Diagram showing the pairwise transition boundaries corresponding to the singular fade state $\sqrt{2}e^{j \pi/4}$ for 4-PSK signal set}	
\label{4psk_regions_example}	
\end{figure}

\begin{figure}[htbp]
\centering
\includegraphics[totalheight=3.5in,width=3.5in]{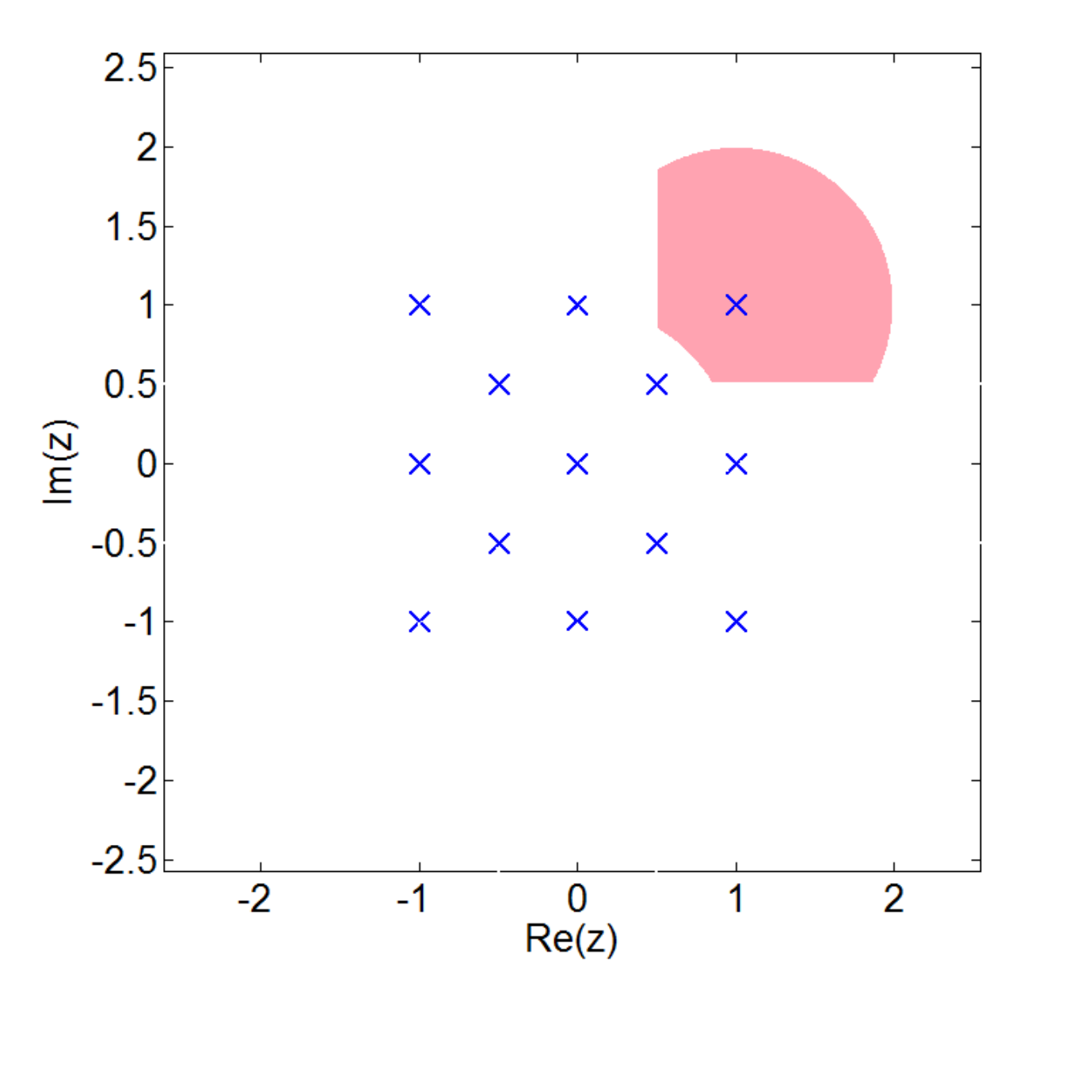}
\caption{Diagram showing the region $\mathcal{R}_{\lbrace\sqrt{2}e^{j \pi/4}\rbrace}$ for 4-PSK signal set}
\label{4psk_regions_ex_shaded}	
\end{figure}

\begin{figure}[htbp]
\centering
\includegraphics[totalheight=3.5in,width=3.5in]{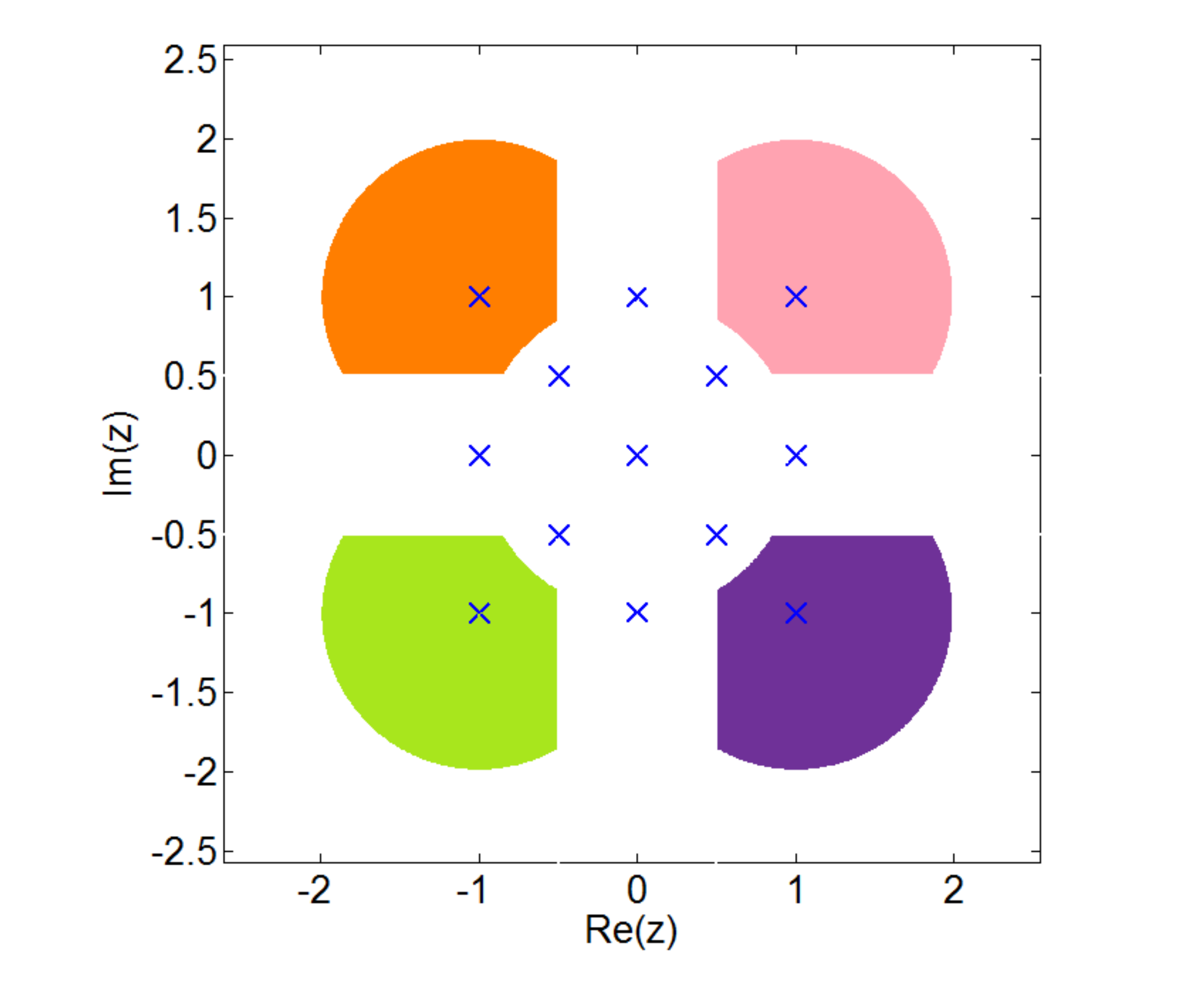}
\caption{Diagram showing the regions corresponding to the singular fade states lying on the circle of radius $\sqrt{2}$ for 4-PSK signal set}	
\label{4psk_ex_45_regions}	
\end{figure}

\begin{figure}[htbp]
\centering
\includegraphics[totalheight=3.5in,width=3.5in]{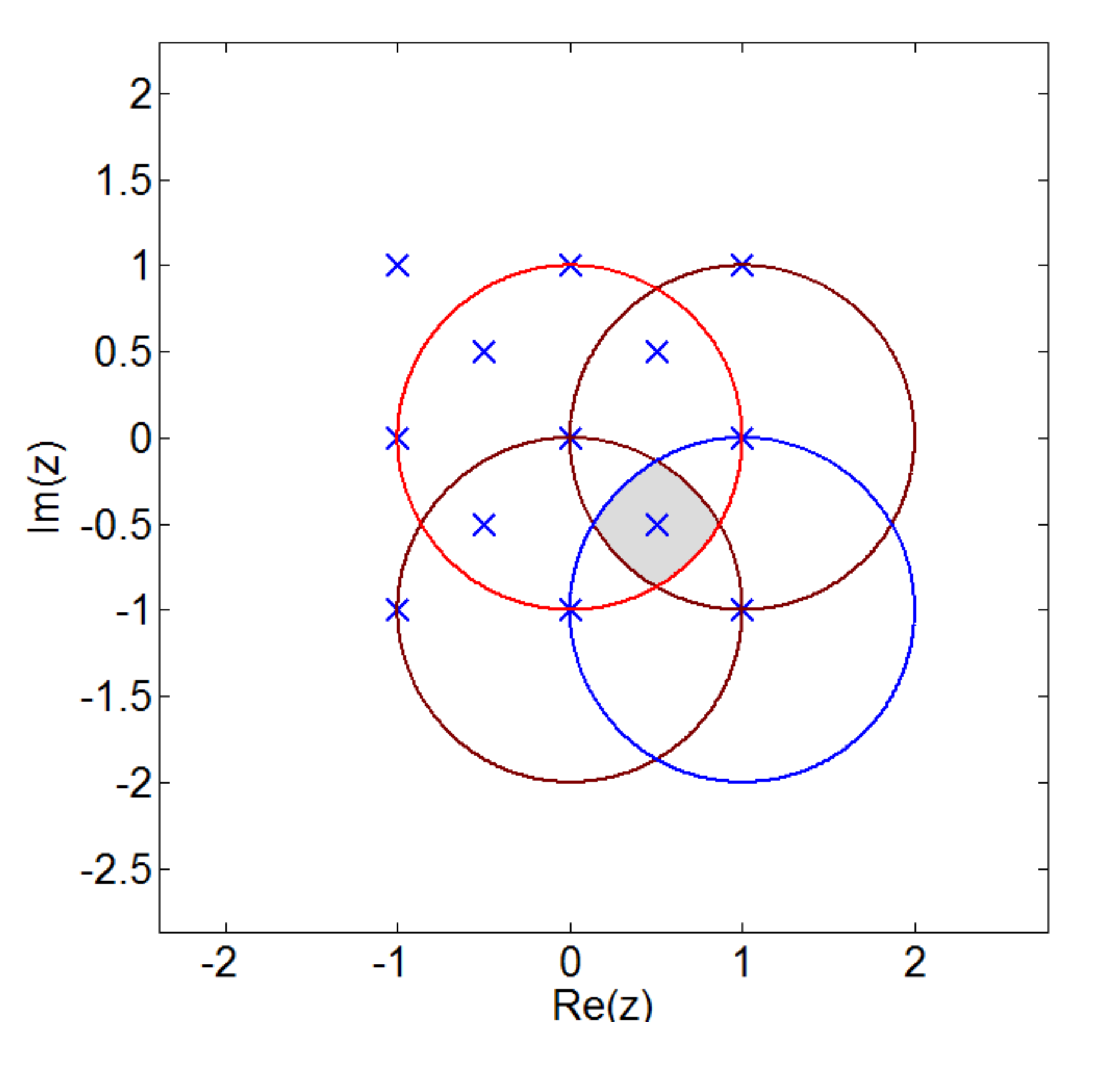}
\caption{Diagram showing the pairwise transition boundaries corresponding to the singular fade state $0.7071e^{-j \pi/4}$ for 4-PSK signal set}	
\label{4psk_region_1byroot2}	
\end{figure}

\begin{figure}[htbp]
\centering
\includegraphics[totalheight=3.5in,width=3.8in]{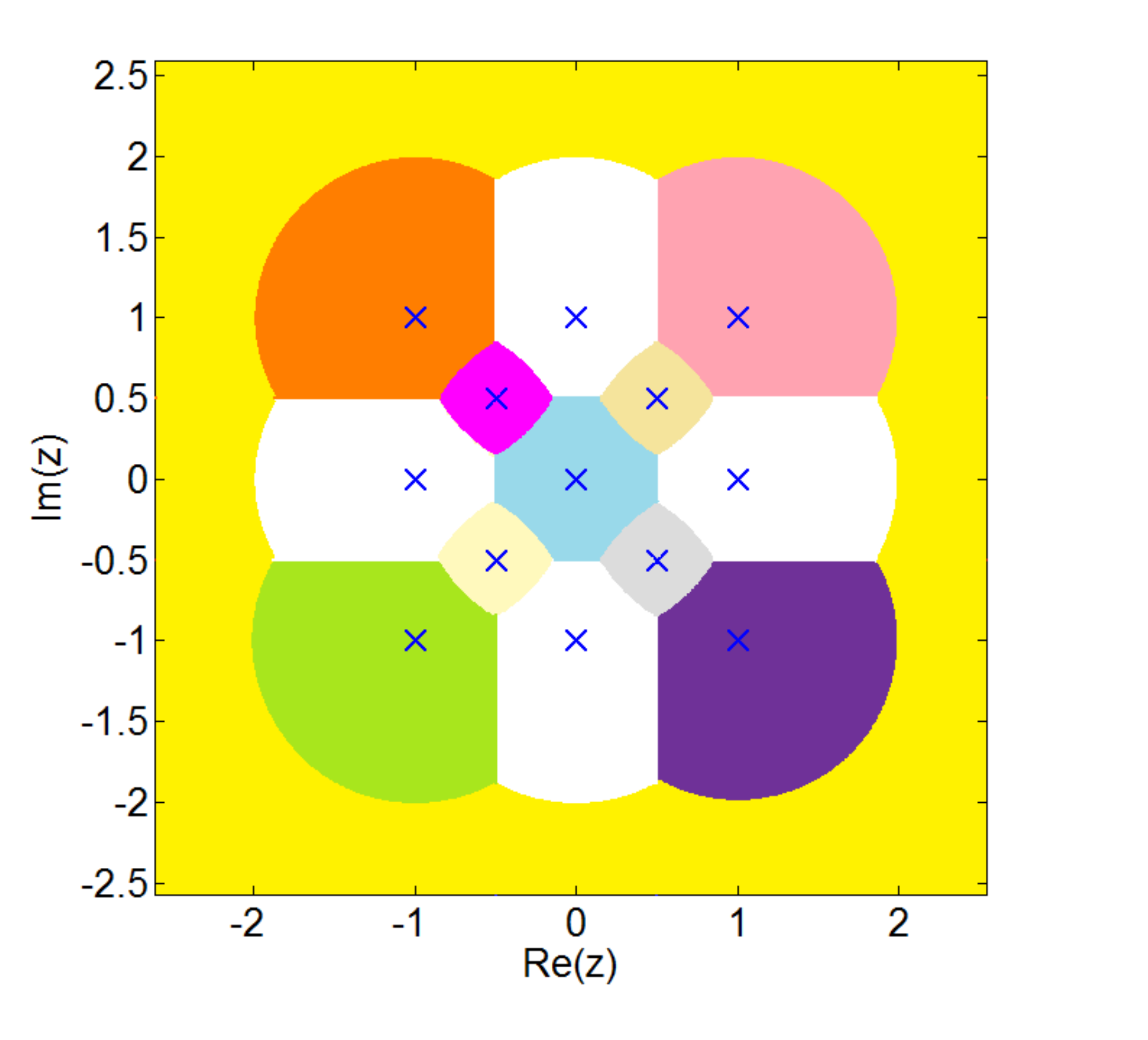}
\caption{Diagram showing the regions associated with the singular fade states on the unit circle}	
\label{4psk_region_all}	
\end{figure}

\begin{figure}[htbp]
\centering
\includegraphics[totalheight=3.5in,width=3.8in]{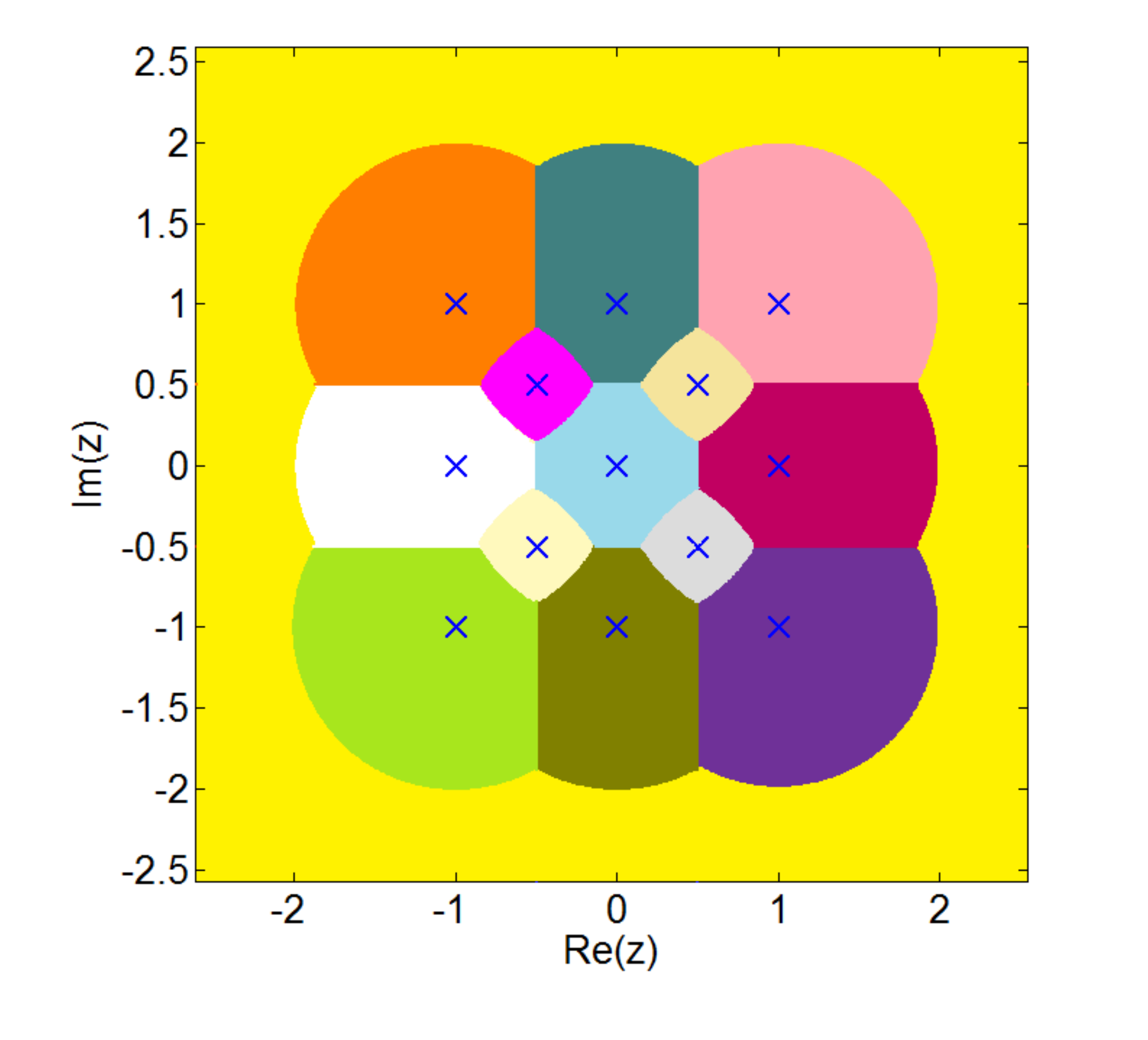}
\caption{Diagram showing the quantization of the $\gamma e^{j \theta}$ plane for 4-PSK signal set}	
\label{4psk_region_all_final}	
\end{figure}

Consider the case when the nodes A and B use 4-PSK signal set. From Lemma \ref{reg_comp_inv}, it follows that it is enough to consider the singular fade states which lie on the circle with radius $\sqrt{2}$. Also, it is enough to consider the region corresponding to the singular fade state $\sqrt{2}e^{j \pi/4}$. The regions corresponding to rest of the singular fade states on the circle with radius $\sqrt{2}$ can be obtained by symmetry.  From Lemma \ref{lemma_wedge}, the region $\mathcal{R}_{\lbrace\sqrt{2}e^{j \pi/4}\rbrace}$ lies inside the wedge formed by the straight lines $\theta=0$ and $\theta=\pi/2$ (shown by green straight lines in Fig. \ref{4psk_regions_example}). To obtain the region $\mathcal{R}_{\lbrace\sqrt{2}e^{j \pi/4}\rbrace}$, from Lemma \ref{lemma:same_line}, we need to consider the unit circle centered at the origin (shown by the red circle in Fig. \ref{4psk_regions_example}). Since the singular fade state $\sqrt{2}e^{j \pi/4}$ falls on the outermost circle, from Lemma \ref{lemma:same_line}, we also need to consider the unit circle centered at $\sqrt{2}e^{j \pi/4}$ (shown by the blue circle in Fig. \ref{4psk_regions_example}). From Lemma \ref{lemma:same_pi_by_m}, we need to consider the curves $c(\sqrt{2}e^{j \pi/4},1)$ and $c(\sqrt{2},e^{j 2\pi/4})$ (shown by  brown straight lines in Fig. \ref{4psk_regions_example}). The region $\mathcal{R}_{\lbrace\sqrt{2}e^{j \pi/4}\rbrace}$, which is the inner envelope region formed by the point $\sqrt{2}e^{j \pi/4}$ and the curves shown in Fig. \ref{4psk_regions_example}, is the shaded region shown in Fig. \ref {4psk_regions_ex_shaded}. The regions corresponding to the other singular fade states which lie on the unit circle with radius $\sqrt{2}$ can be obtained by symmetry and are shown in Fig. \ref{4psk_ex_45_regions}. 

The curves in Fig. \ref{4psk_regions_example}, after complex inversion, become the curves shown in Fig. \ref{4psk_region_1byroot2}. By Lemma 5, taking the inner envelope region of the curves in Fig. \ref{4psk_region_1byroot2}, we get the region $\mathcal{R}_{\lbrace 0.7071 e^{-j \pi/4} \rbrace}$ (the region shaded grey in Fig. \ref{4psk_region_1byroot2}). The regions corresponding to the other singular fade states which lie on the circle with radius $0.7071$ obtained by symmetry are shown in Fig. \ref{4psk_region_all}.  Fig. \ref{4psk_region_all} also shows singularity-free region (the regions shaded yellow and blue) which were obtained in Section III. The remaining portion of the singularity region (left unshaded in Fig. \ref{4psk_region_all}) are the regions associated with the singular fade states which lie on the unit circle. Putting all the pieces together, the channel quantization for 4-PSK signal set is as shown in Fig. \ref{4psk_region_all_final}.

The channel quantization for 4-PSK signal set shown in Fig. \ref{4psk_region_all_final} is same as the one obtained by computer search in  Fig. 11 in \cite{KoPoTa}. 
\section{SIMULATION RESULTS}
\begin{figure}[htbp]
\centering
\includegraphics[totalheight=2.75in,width=3.5in]{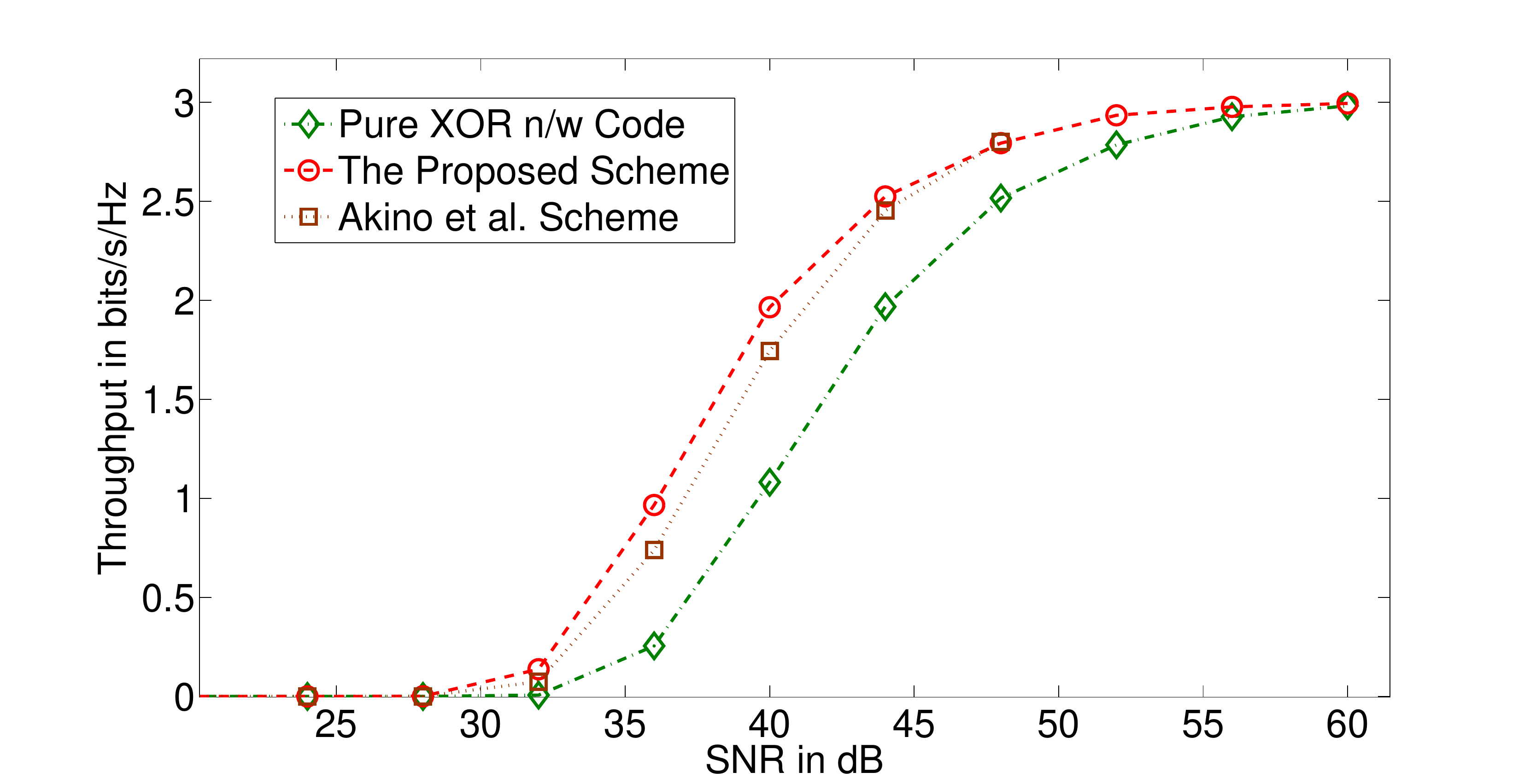}
\caption{SNR vs Throughput for different schemes for 8-PSK signal set}	
\label{tput_curves}	
\end{figure}

 It is assumed that $H_A$, $H_B$, $H'_A$ and $H'_B$ are distributed according to Rayleigh distribution. The variances of all the fading links are assumed to be 0 dB. A frame length of 256 symbols is assumed for each transmission. As mentioned earlier, for the case when $4$-PSK signal set is used at the end nodes, the channel quantization obtained reduces to the one obtained using Algorithm 2 provided in \cite{KoPoTa} (for simulation results for this case, see \cite{KoPoTa}). The throughput in bits/s/Hz as a function of average SNR for the proposed scheme is shown in  Fig. \ref{tput_curves}, for the case when 8-PSK signal set is used at the nodes during the MA phase. Fig. \ref{tput_curves} also shows the throughput vs SNR curves for the scheme based on computer search proposed by Koike-Akino et. al (Algorithm 1, \cite{KoPoTa}) and for the case when pure-XOR network code \cite{KoPoTa} is used irrespective of the channel condition. For the proposed scheme, the set of clusterings which remove the singular fade states obtained in \cite{LS} are considered and for a particular realization of channel fade state, the criterion given in Section IV A is used to select a clustering. The proposed scheme and the scheme suggested by Koike-Akino et al. outperform the scheme which uses only XOR network code, since all the singular fade states are not removed by the XOR network code. From Fig. \ref{tput_curves}, it can be seen that the proposed scheme outperforms Koike-Akino et al. scheme. The reason for this is that the proposed scheme uses 8-point signal set during the BC phase under all channel conditions, while Koike-Akino et al. scheme uses a signal set of cardinality ranging from 8 to 14 depending on the channel condition. 
\section{DISCUSSION}
  The design of modulation schemes for the physical layer network-coded two way relaying scenario was considered. It was shown that the set of possible channel realizations (the complex plane) can be broadly classified in to two regions: the singularity-free region and the singularity region. In the singularity-free region, it was shown shown that any clustering satisfying the exclusive law gives the same minimum cluster distance. The singularity-free region was obtained analytically for M-PSK signal set. A partition of the singularity region was obtained based on the criteria of removing the singular fade states. Throughout, it was assumed that, there exist clusterings which remove the singular fade states. Such clusterings, which remove the singular fade states, can be obtained from \cite{KoPoTa}, where a computer search algorithm is provided, or from \cite{LS}, where an analytical approach based on Latin squares is presented.   
\section*{Acknowledgement}
This work was supported  partly by the DRDO-IISc program on Advanced Research in Mathematical Engineering through a research grant as well as the INAE Chair Professorship grant to B.~S.~Rajan.

\appendices
\section{Proof of Lemma 1}
\begin{proof}
The if part of the lemma is straight forward. The proof of the only if part is by contradiction. Assume $k_1 \neq l_1$ and $k_2 \neq l_2$. Since $M$ is a power of 2, let $M=2^{\lambda-1}$, where $\lambda \geq 2$ is an integer. Consider the extension field $\mathbb{K}= \mathbb{Q}(e^{j 2 \pi/2^\lambda })$ of $\mathbb{Q}$, which is the smallest field containing $\mathbb{Q}$ and $e^{j 2 \pi/2^\lambda }$. The field $\mathbb{K}$ forms a vector space of dimension $\varphi(2^\lambda)=2^{\lambda-1}$ over $\mathbb{Q}$, where $\varphi$ denotes Euler's totient function. The set $\lbrace 1, e^{j \frac{2\pi}{2^\lambda}}, e^{j  \frac{2\pi}{2^\lambda}2},...,e^{j \frac{2\pi}{2^\lambda}(2^{\lambda-1}-1)}\rbrace$, forms a set of basis vectors for the vector space. Assume that $\dfrac{\sin(k_1 \pi/M)}{\sin(k_2 \pi/M)}=\dfrac{\sin(l_1 \pi/M)}{\sin(l_2 \pi/M)}$, which is the same as,

{\footnotesize
\begin{align}
\nonumber
e^{j \frac{2\pi}{2^{\lambda}}(k_1+l_2)}+e^{-j \frac{2\pi}{2^{\lambda}}(k_1+l_2)}-e^{j \frac{2\pi}{2^{\lambda}}(k_1-l_2)}-e^{-j \frac{2\pi}{2^{\lambda}}(k_1-l_2)}=\\
\label{eqn1_main}
e^{j \frac{2\pi}{2^{\lambda}}(k_2+l_1)}+e^{-j \frac{2\pi}{2^{\lambda}}(k_2+l_1)}-e^{j \frac{2\pi}{2^{\lambda}}(k_2-l_1)}-e^{-j \frac{2\pi}{2^{\lambda}}(k_2-l_1)}.
\end{align}
}

The quantities on the L.H.S and the R.H.S of \eqref{eqn1_main} are elements of the field $\mathbb{K}$ expressed as a linear combination of the basis vectors. Since this representation is unique, we can equate the set of complex exponentials which appear on the L.H.S and R.H.S of \eqref{eqn1_main} and the corresponding scalars which scale these complex exponentials need to be equal. As a result, we have, $k_1+l_2=\pm (k_2+l_1)$ and $k_1-l_2 = \pm (k_2-l_1)$. The above two equations imply either $k_1=k_2, l_1=l_2$ or $k_1=l_1,k_2=l_2$, resulting in a contradiction. This completes the proof.
\end{proof}
\section{}
Before we prove Theorem 1, we introduce some notations and prove some lemmas. 

Let $\Gamma_{SF}$ denote the singularity-free region defined in Section III. Let $\Gamma ^{ext}_{SF}$ denote the region $ \Gamma _{SF} \cap \lbrace \gamma >0\rbrace$, i.e., 

{\footnotesize
\begin{align*}
\Gamma ^{ext}_{SF}=\lbrace \gamma e^{j \theta} : \vert d_k+\gamma e^{j \theta}d_l\vert \geq \min(2 \sin(\pi/M),2\gamma \sin(\pi/M)), \\
\forall d_k,d_l \in \Delta\mathcal{S}-, \gamma >1 , -\pi \leq \theta < \pi\rbrace.
\end{align*}
}

Since, for $\gamma >1$, $\min(2 \sin(\pi/M),2\gamma \sin(\pi/M))= 2 \sin(\pi/M)$,

{\footnotesize
\begin{align}
\nonumber
\Gamma ^{ext}_{SF}=\lbrace \gamma e^{j \theta} &: \vert d_k+\gamma e^{j \theta}d_l\vert \geq 2 \sin(\pi/M), \\
\label{region_ext}
&\hspace{1cm}\forall d_k,d_l \in \Delta\mathcal{S}-, \gamma >1 , -\pi \leq \theta < \pi\rbrace.
\end{align}
}

Similarly, the region $\Gamma ^{int}_{SF}$  is defined to be $ \Gamma _{SF} \cap \lbrace \gamma \leq 0\rbrace$, i.e.,

{\footnotesize
\begin{align}
\nonumber
\Gamma ^{int}_{SF}=\lbrace \gamma e^{j \theta} &: \vert d_k+\gamma e^{j \theta}d_l\vert \geq 2\gamma \sin(\pi/M), \\
\label{region_int}
&\hspace{1cm}\forall d_k,d_l \in \Delta\mathcal{S}, \gamma \leq 1 , -\pi \leq \theta < \pi\rbrace.
\end{align}
}

\begin{lemma}
\label{sf_int_ext}
The region $\Gamma^{int}_{SF}$ is obtained by the complex inversion of the region $\Gamma^{ext}_{SF}$.
\begin{proof}
Using the transformation $\gamma'e^{j \theta'}=\dfrac{1}{\gamma e^{j \theta}}$ in \eqref{region_ext}, we get the region,

{\footnotesize
\begin{align*}
\lbrace \gamma' e^{j \theta'} : \vert d_k+d_l/(\gamma' e^{j \theta'})\vert &\geq 2 \sin(\pi/M), \forall d_k,d_l \in \Delta\mathcal{S},\\
 &\hspace{2.5cm}\gamma' <1 , -\pi \leq \theta < \pi\rbrace\\
 =\lbrace \gamma' e^{j \theta'} : \vert \gamma' e^{j \theta'}d_k+d_l\vert &\geq 2 \gamma' \sin(\pi/M), \forall d_k,d_l \in \Delta\mathcal{S},\\
 &\hspace{2.5cm}\gamma' <1 , -\pi \leq \theta' < \pi\rbrace,
\end{align*}
}which is the same as $\Gamma^{int}_{SF}$.
 
\end{proof}
\end{lemma}

From Lemma \ref{sf_int_ext}, it is clear that once we obtain $\Gamma^{ext}_{SF}$, $\Gamma^{int}_{SF}$ can be obtained by complex inversion. Hence, in the following discussion it is assumed that $\gamma >1$.

Let $c_{k_1,k_2},1 \leq k_1,k_2 \leq M/2, k_1 > k_2$ denote the circles centered at the origin with radii $\sin(\pi k_1/M)/\sin(\pi k_2/M)$. Let $c_{1,1}$ denote the unit circle centered at the origin. Let $C_{k_1,k_2}, \leq k_1,k_2 \leq M/2, k_1 \neq k_2$ denote the set of circles whose centers are the singular fade states which lie on $c_{k_1,k_2}$ and radii equal to  $\sin(\pi/M)/\sin( k_2\pi/M)$. Let $C_{1,1}$ denote the set of circles whose centers are the singular fade states which lie on $c_{1,1}$ and have unit radii.

\begin{lemma}
The region $\Gamma^{ext}_{SF}$ is the unshaded region obtained when the interior regions of all the circles which belong to the sets $C_{k_1,k_2}, \leq k_1,k_2 \leq M/2, k_1 \neq k_2$ are shaded.
\begin{proof}
From Section II, it can be seen that the points in the difference constellation $\Delta\mathcal{S}$ lie on circles with radius $2 \sin(n \pi/M)$, for some $1 \leq n \leq M/2$. The region $\Gamma^{ext}_{SF}$ is given in \eqref{region_ext}. Let $\vert d_k\vert= 2 \sin(\pi k_1/M)$ and $\vert d_l\vert= 2 \sin(\pi k_2/M)$. From \eqref{region_ext}, we have

{\footnotesize
\begin{align*}
\Gamma ^{ext}_{SF}=\lbrace \gamma e^{j \theta} &: \vert d_k/d_l+\gamma e^{j \theta}\vert \geq \sin(\pi/M)/\sin(\pi k_2/M), \\
&\hspace{1cm}\forall d_k,d_l \in \Delta\mathcal{S}, \gamma >1 , -\pi \leq \theta < \pi\rbrace.
\end{align*}
}

The result follows from the fact that $\vert d_k/d_l+\gamma e^{j \theta}\vert < \sin(\pi/M)/\sin(\pi k_2/M)$ is the interior of the circle with center at the singular fade state $-d_k/d_l$ and radius $\sin(\pi/M)/\sin(\pi k_2/M)$.

\end{proof}
\end{lemma} 
In the rest of the discussion, it is imagined that the interiors of all the circles which belong to the sets $C_{k_1,k_2}, \leq k_1,k_2 \leq M/2, k_1 \neq k_2$ are shaded.

The circles $c_{n,n'},1 \leq n' \leq n, 1 \leq n \leq M/2$ are split in to different groups as follows: $g_i=\lbrace c_{i+1,i},c_{i+2,i},...c_{M/2,i}\rbrace,1 \leq i \leq M/2$. The region between the outermost circle and the innermost circle in $g_i$ is called the ring formed by $g_i$. Note that the outermost and innermost circles in $g_i$ are respectively $c_{M/2,i}$ and $c_{i+1,i}$.

\begin{lemma}
\label{trig_ineq}
The inequality,

{\footnotesize
\begin{align} 
\label{eqn_lemma}
 \cos\left(\dfrac{(2k+1) \pi}{2M}\right)\sin\left(\dfrac{\pi}{2M}\right) < \cos\left(\dfrac{(k+1) \pi}{M}\right)\sin\left(\dfrac{\pi}{M}\right),
\end{align}
} holds for $1 \leq k \leq M/2-2$ and is reversed for $k=M/2-1$.
\begin{proof}
We have,  for $1 \leq k<M/2$,

{\footnotesize
\begin{align}
 \label{lemma:eqn1} 
\cos\left(\dfrac{(2k+1) \pi}{2M}\right)\sin\left(\dfrac{\pi}{2M}\right)  < \cos\left(\dfrac{(2k-1) \pi}{2M}\right)\sin\left(\dfrac{\pi}{2M}\right),
\end{align}
}since $\cos x$ is a decreasing function of $x$ for $0 \leq x \leq \pi/2$.
 For $1 \leq k \leq M/2-2$, since $\ 0<(2k+3){\pi}/{2M} < \pi/2$,
 {\footnotesize
 \begin{align}
 \label{lemma:eqn2} 
 \cos\left(\dfrac{(2k+3)\pi}{2M}\right)>0.
 \end{align}
 }
 
  Hence, from \eqref{lemma:eqn1} and \eqref{lemma:eqn2} we have,  for $1 \leq   k \leq M/2-2$,
  
  {\footnotesize
  \begin{align}
  \nonumber
  \cos\left(\dfrac{(2k+1) \pi}{2M}\right)\sin\left(\dfrac{\pi}{2M}\right) & \\
  \label{lemma:eqn3}
  &\hspace{-2.3 cm}< \cos\left(\dfrac{(2k-1) \pi}{2M}\right)\sin\left(\dfrac{\pi}{2M}\right)+\cos\left(\dfrac{(2k+3)\pi}{2M}\right)\sin\left(\dfrac{\pi}{2M}\right)
  \end{align}
  } 
  
  From \eqref{lemma:eqn3}, using standard trigonometric identities, it follows that,
  {\footnotesize
  \begin{align}
  \nonumber
  \cos\left(\dfrac{(2k+1) \pi}{2M}\right)\sin\left(\dfrac{\pi}{2M}\right) 
  &  < 2 \sin\left(\dfrac{\pi}{2M}\right) \cos\left(\dfrac{\pi}{2M}\right)\cos\left(\dfrac{(k+1)\pi}{M}\right)\\
  \nonumber
  &=\cos\left(\dfrac{(k+1) \pi}{M}\right)\sin\left(\dfrac{\pi}{M}\right).
  \end{align}
  }
  
  For $k=M/2-1$, the right hand side of the inequality \eqref{eqn_lemma} is zero whereas the left hand side is a positive quantity and hence the inequality \eqref{eqn_lemma} reverses direction.
\end{proof} 
\end{lemma}

\begin{lemma}
\label{lemma_ring}
The rings formed by $g_i, 1 \leq i \leq M/2$ are fully shaded.
\begin{proof}
To prove the lemma, we show that the region between the circles $c_{k,i}$ and $c_{k+1,i}$ is fully shaded, $\forall 1 \leq i \leq M/2-1,i+1 \leq k \leq M/2$. The circles $c_{k,i}$ and $c_{k+1,i}$ shown in Fig. \ref{fig:circle} have centers at the origin $O$. Let $A$ and $B$ be two points on the circle $c_{k+1,i}$ with angular separation $2\pi/M$, which are the centers of two  circles in the set $C_{k+1,i}$, whose intersection points are denoted as $C$ and $D$ in Fig. \ref{fig:circle}. Let $E$ and $F$ denote the intersection points of the line segment $CD$ with the circles $c_{k+1,i}$ and $c_{k,i}$ respectively. From Fig. \ref{fig:circle}, to show that the region between the circles $c_{k,i}$ and $c_{k+1,i}$ is completely shaded, it is enough to show that $EF =r_{k+1,i}-r_{k,i}\leq EC$ and $EC \leq DC$, where $r_{k+1,i}$ and $r_{k,i}$ are the radii of the circles $c_{k+1,i}$ and $c_{k,i}$ respectively.

From Fig. \ref{fig:circle}, it can be seen that $AB/2=\sin(\angle DOB )r_{k+1,i}=\sin(\pi/M)r_{k+1,i}=\sin(\angle DCB )BC$. The length of the diagonal of the rhombus $ADBC$ can be shown to be 

{\footnotesize
\begin{align}
\nonumber
DC=BC\sqrt{1-\cos(2\angle DCB)}&=2\sqrt{BC^2-r_{k+1,i}^2\sin^2(\pi/M)}\\
 \label{eqn_DC}
&=2\left[\dfrac{\sin\left(\dfrac{\pi}{M}\right)\cos\left(\dfrac{(k+1)\pi}{M}\right)}{\sin\left(\dfrac{i\pi}{M}\right)}\right].
\end{align}
}


The distance, 
{\footnotesize
\begin{align}
\nonumber
OC&=OG-CG=OG-DC/2\\
\nonumber
&=\dfrac{\sin\left(\dfrac{(k+1)\pi}{M}\right)\cos(\pi/M)}{\sin\left(\dfrac{i \pi}{M}\right)}-\dfrac{\sin\left(\dfrac{\pi}{M}\right)\cos\left(\dfrac{(k+1)\pi}{M}\right)}{\sin\left(\dfrac{i\pi}{M}\right)}\\
\label{eqn_OC}
&=\dfrac{\sin\left(\dfrac{k\pi}{M}\right)}{\sin\left(\dfrac{i\pi}{M}\right)}.
\end{align}
}
From \eqref{eqn_OC}, it can be seen that $OC=OF$, i.e., the points $C$ and $F$ exactly coincide. Hence $EC=EF$.  The distance,

{\footnotesize
\begin{align} 
\nonumber
 EC=EF&=\dfrac{\sin\left(\dfrac{\left(k+1\right)\pi}{M}\right)-\sin\left(\dfrac{k\pi}{M}\right)}{\sin\left(\dfrac{i\pi}{M}\right)}\\
 \label{eqn_EC}
 &=2\dfrac{\cos\left(\dfrac{(2k+1) \pi}{2M}\right)\sin\left(\dfrac{\pi}{2M}\right)}{\sin\left(\dfrac{i\pi}{M}\right)}.
 \end{align}
 }  
 From \eqref{eqn_DC}, \eqref{eqn_EC} and Lemma \eqref{trig_ineq}, it follows that $EC \leq DC$.
\begin{figure}[htbp]
\centering
\vspace{-3 cm}
\includegraphics[totalheight=6in,width=4in]{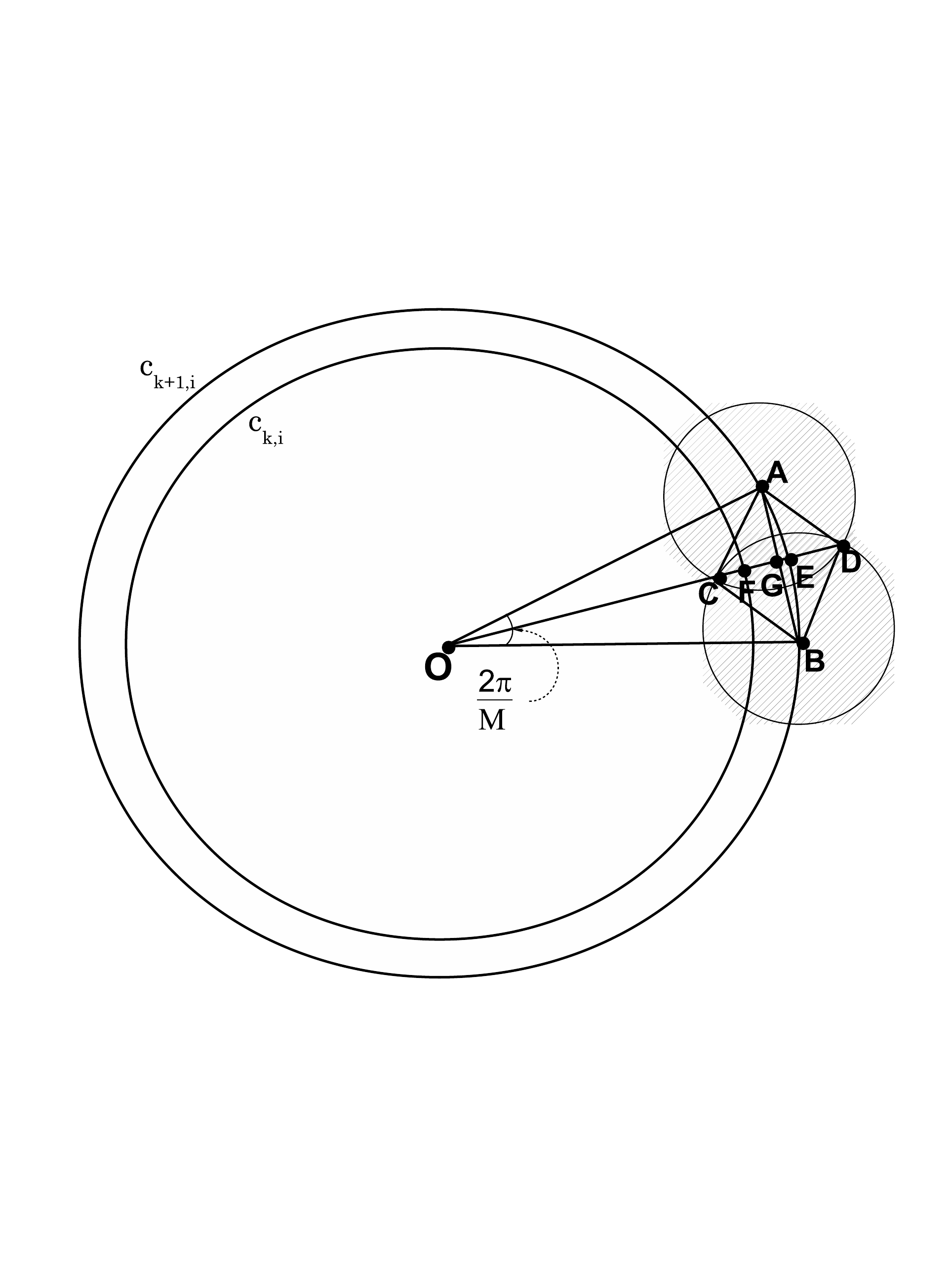}
\vspace{-3 cm}
\caption{Diagram showing the circles used in the proof of Lemma \ref{lemma_ring}}	
\label{fig:circle}	
\end{figure}
\end{proof}
\end{lemma}

Among all the circles $c_{k_1,k_2}, k_1 \neq k_2$, $c_{M/2,1}$ is the outermost. In the following lemma, it is shown that the region between the circles $c_{M/2,1}$ and $c_{1,1}$ is fully shaded.
\begin{lemma}
\label{fully_shaded}
The region between the circles $c_{M/2,1}$ and $c_{1,1}$ is fully shaded.

\begin{proof}
In between $c_{M/2,1}$ and $c_{1,1}$, the circles $c_{k_1,k_2}, k_1 \neq k_2$ form different rings which are fully shaded. A pair of rings $g_i, 1 \leq i \leq M/2-1$ may overlap or may not overlap. In either case, to prove the lemma, it is enough to show that the region between the innermost circle in $g_i$ (which is $c_{i+1,i}$) and $c_{1,1}$  is fully shaded, $\forall 1 \leq i \leq M/2-1$.

The circles $c_{1,1}$ and $c_{i+1,i}$ are shown in Fig. \ref{fig:circles2}. Also shown are two shaded circles which belong to the set $C_{1,1}$, whose centers $A$ and $B$ differ by an angular separation of $2\pi/M$ and have unit radius. Since $OC$ is the bisector of $\angle EOD$, we have $\angle COD = \pi/M$. Also, $\angle ODC=\pi/2$ (angle in a semi-circle). Hence, we have $OC=OD \cos(\pi/M)= 2 \cos(\pi/M)$. To show that the region between $c_{i+1,i}$ and $c_{1,1}$ is completely shaded, it is enough to show that the radius of $c_{i+1,i}$ is less than $OC$, i.e., we need to show,

{\footnotesize
\begin{align}
\label{eqn_lemma_final}
 \dfrac{\sin\left(\dfrac{(i+1)\pi}{M}\right)}{\sin\left(\dfrac{i \pi}{M}\right)}\leq 2 \cos\left({\pi}{M}\right).
\end{align}
 }
  Since $\cot(x)$
  is a decreasing function of  $x$, we have, $\cot(i \pi /M) \leq \cot(\pi/M)$, i.e.,
  
{\footnotesize
\begin{align}  
\label{eqn_temp}   
 \dfrac{ \sin\left(\dfrac{\pi}{M}\right) \cos\left(\dfrac{i \pi}{M}\right)}{\sin\left(\dfrac{i \pi}{M}\right)} \leq \cos\left(\dfrac{\pi}{M}\right). 
  \end{align}
 }
  
  Adding $\cos(\pi/M)$ to the both the sides of \eqref{eqn_temp} we get, 
  
  {\footnotesize
  \begin{align*}
  \dfrac{\sin\left(\dfrac{\pi}{M}\right) \cos\left(\dfrac{i \pi}{M}\right)+\cos\left(\dfrac{\pi}{M}\right) \sin\left(\dfrac{i \pi}{M}\right)}{\sin\left(\dfrac{i \pi}{M}\right)} \leq 2\cos\left(\dfrac{\pi}{M}\right),
  \end{align*}
  }which is the same as \eqref{eqn_lemma_final}.
\begin{figure}[htbp]
\centering
\vspace{-1 cm}
\includegraphics[totalheight=3.5in,width=3.5in]{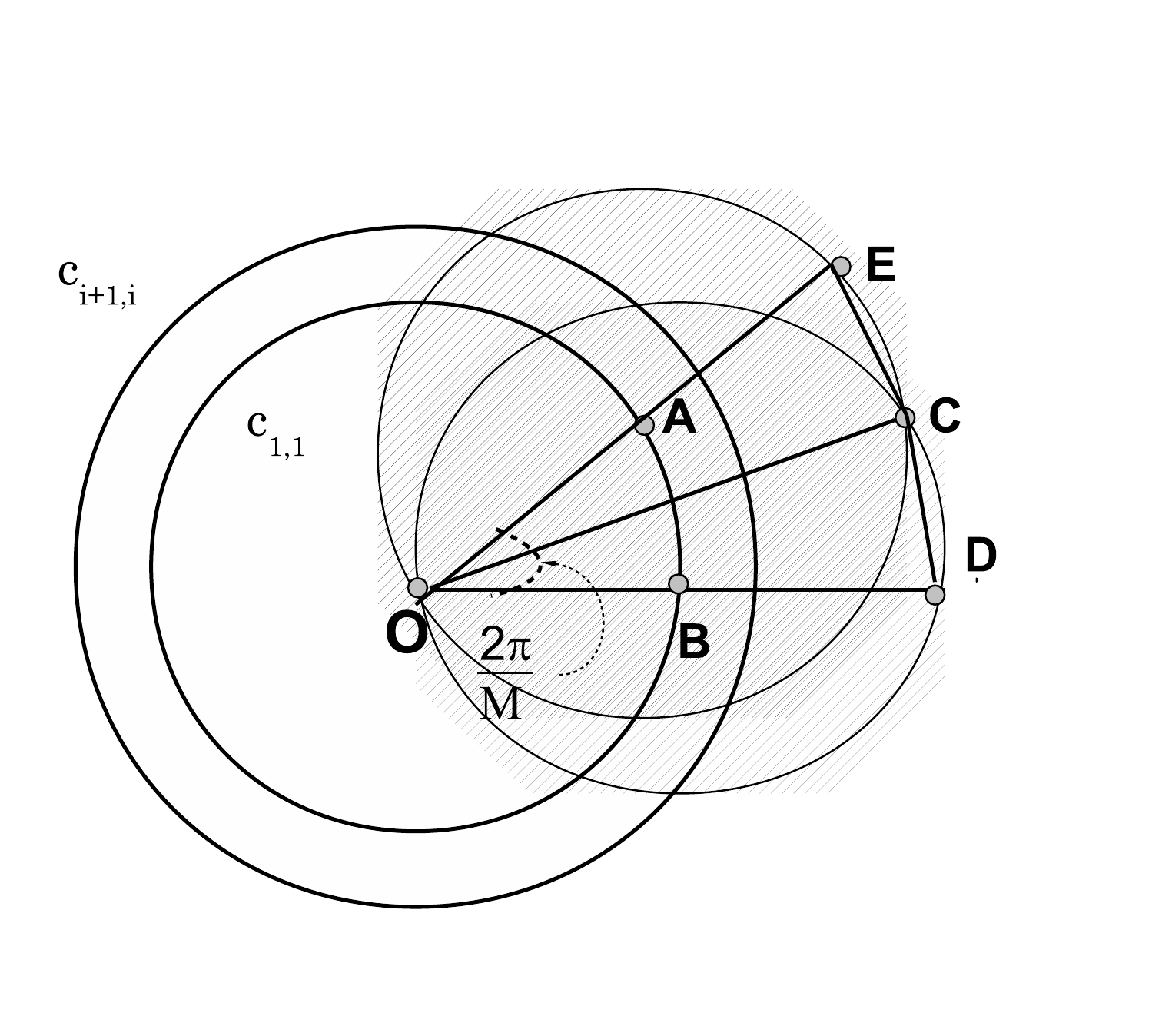}
\vspace{-1 cm}
\caption{Diagram showing the circles used in the proof of Lemma \ref{fully_shaded}}	
\label{fig:circles2}	
\end{figure}
\end{proof}
\end{lemma}

\begin{lemma}
\label{lemma_tot_inside}
For $M>4$, all the circles which belong to the sets $C_{k_1,k_2}, k_1 \neq k_2 , 1 \leq k_1 \leq M/2$ and $2 \leq k_2 \leq M/2$ totally lie inside the circle $c_{M/2,1}$ , 

\begin{proof}
 The circles belonging to the set $C_{k_1,k_2}$ have the absolute value of their centers $\sin(k_1 \pi/M)/\sin(k_2 \pi /M)$ and radius $\sin(\pi/M)/\sin(k_2 \pi /M)$ . The circle $c_{M/2,1}$ is centered at the origin and has radius $1/\sin(\pi/M)$. Hence, to prove the Lemma, it is enough to show that  for $M>4$
 
{\footnotesize 
\begin{align*} 
 \dfrac{\sin\left(\dfrac{k_1 \pi}{M}\right)}{\sin\left(\dfrac{k_2 \pi} {M}\right)}+\dfrac{\sin\left(\dfrac{\pi}{M}\right)}{\sin\left(\dfrac{k_2 \pi} {M}\right)} \leq \dfrac{1}{\sin\left(\dfrac{\pi}{M}\right)}, 1 \leq k_1 \leq \dfrac{M}{2},2 \leq k_2 \leq \dfrac{M}{2}.
 \end{align*}
 }
 It is enough to show that,
  {\footnotesize 
\begin{align*} 
  \max _{1 \leq k_1 \leq M/2} \left[ \sin\left(\dfrac{k_1 \pi} {M}\right) + \sin\left(\dfrac{\pi}{M}\right) \right] \leq \min_{ 2 \leq k_2 \leq M/2} \left[\dfrac{\sin\left(\dfrac{k_2 \pi}{M}\right)}{\sin\left(\dfrac{\pi}{M}\right)}\right],
  \end{align*}
  } i.e., we need to show that $1+\sin(\pi/M) \leq 2 \cos(\pi/M)$, for $M>4$. Let $x'$ be the solution of the equation $1+\sin(\pi/x) = 2 \cos(\pi/x)$. The function $1+ \sin(\pi/x)$ is a monotonically decreasing function of $x$ and $2 \cos(\pi/x)$ is a monotonically increasing function of $x$.  Hence, for $M \geq \lceil x' \rceil$, where $\lceil x' \rceil$ is the least integer greater than or equal to $x'$, $1+\sin(\pi/M) \leq 2 \cos(\pi/M)$. It can be verified that $x'=\dfrac{\pi}{\arcsin(3/5)}$ and hence $\lceil x' \rceil = 5$. 
 \end{proof}
 \end{lemma}

\begin{lemma}
\label{lemma_outer_two}
Consider the circles which belong to the sets $C_{k_1,1}$, $k_1 \leq M/2-2$. The portion of the interior region (of the circles considered) which lies outside the circle $c_{M/2,1}$, lies  in the interior region of the set of circles $C_{M/2,1}$, if $k_1$ is even and in the interior region of the set of circles $C_{M/2-1,1}$, if $k_1$ is odd.
\begin{proof}

We prove the lemma for the case when $k_1$ is even. The proof for the case when $k_1$ is odd is similar and is omitted.
The circles which belong to the set $C_{M/2,1}$ are of unit radii and their centers lie on the circle $c_{M/2,1}$. The circles which belong to the sets $C_{k_1,1}$, $k_1 \leq M/2-2$, $k_1$ even, also have unit radii and have  their centers  on the circle $c_{k_1,1}$. Since $k_1$ is even, the phase angles of the centers of the circles in the sets $C_{M/2,1}$ and $C_{k_1,1}$ are the same. Two circles belonging to the set $C_{M/2,1}$, whose centers have phase angles $0$ and $2 \pi/M$ are shown in Fig. \ref{circles3}. Also shown in Fig. \ref{circles3} is the circle in the set $C_{k_1,1}$ whose center has a phase angle $0$. Following a procedure similar to the one used in Lemma \ref{lemma_ring}, it can be shown that the circles in the set $C_{M/2,1}$ touch each other at a point denoted by $C$ (see Fig. \ref{circles3}), the distance $OC=\cos(\pi/M)/\sin(\pi/M)$ and the angle $\angle COB= \pi/M$.  It is enough to show that $DC \geq DE=1$. The distance $CD= \sqrt{OC^2+OD^2-2OC \:OD \cos(\pi/M)}$. Hence, we have,

{\footnotesize
\begin{align}
\nonumber
 CD^2\triangleq f(k_1)=\dfrac{\cos^2\left(\dfrac{\pi}{M}\right)+\sin^2\left(\dfrac{k_1\pi}{M}\right)-2\cos^2\left(\dfrac{\pi}{M}\right)\sin\left(\dfrac{k_1\pi}{M}\right)}{\sin^2\left(\dfrac{\pi}{M}\right)}.
 \end{align}
 }
 
 We have,
 
 {\footnotesize
 \begin{align}
 \nonumber
 f'(k_1)=\dfrac{2\pi\cos\left(\dfrac{k_1\pi}{M}\right)\left(\sin\left(\dfrac{k_1 \pi}{M}\right)-\cos^2\left(\dfrac{\pi}{M}\right) \right)}{M \sin^2\left(\dfrac{\pi}{M}\right)}.
 \end{align}
} 
 Since $\sin(k_1\pi/M) < \cos^2(\pi/M)$ for $2 \leq k_1 \leq M/2-2$ $f'(k_1)<0$ and hence $f(k_1)$ is a decreasing function of $k_1$. Hence to show that $f(k_1) \geq 1 , \forall 2 \leq k_1 \leq M/2-2$, it is enough to show that $f(M/2-2) \geq 1$. It can be verified that $f(M/2-2)=1$. 
\begin{figure}[htbp]
\centering
\vspace{-1 cm}
\includegraphics[totalheight=6in,width=4in]{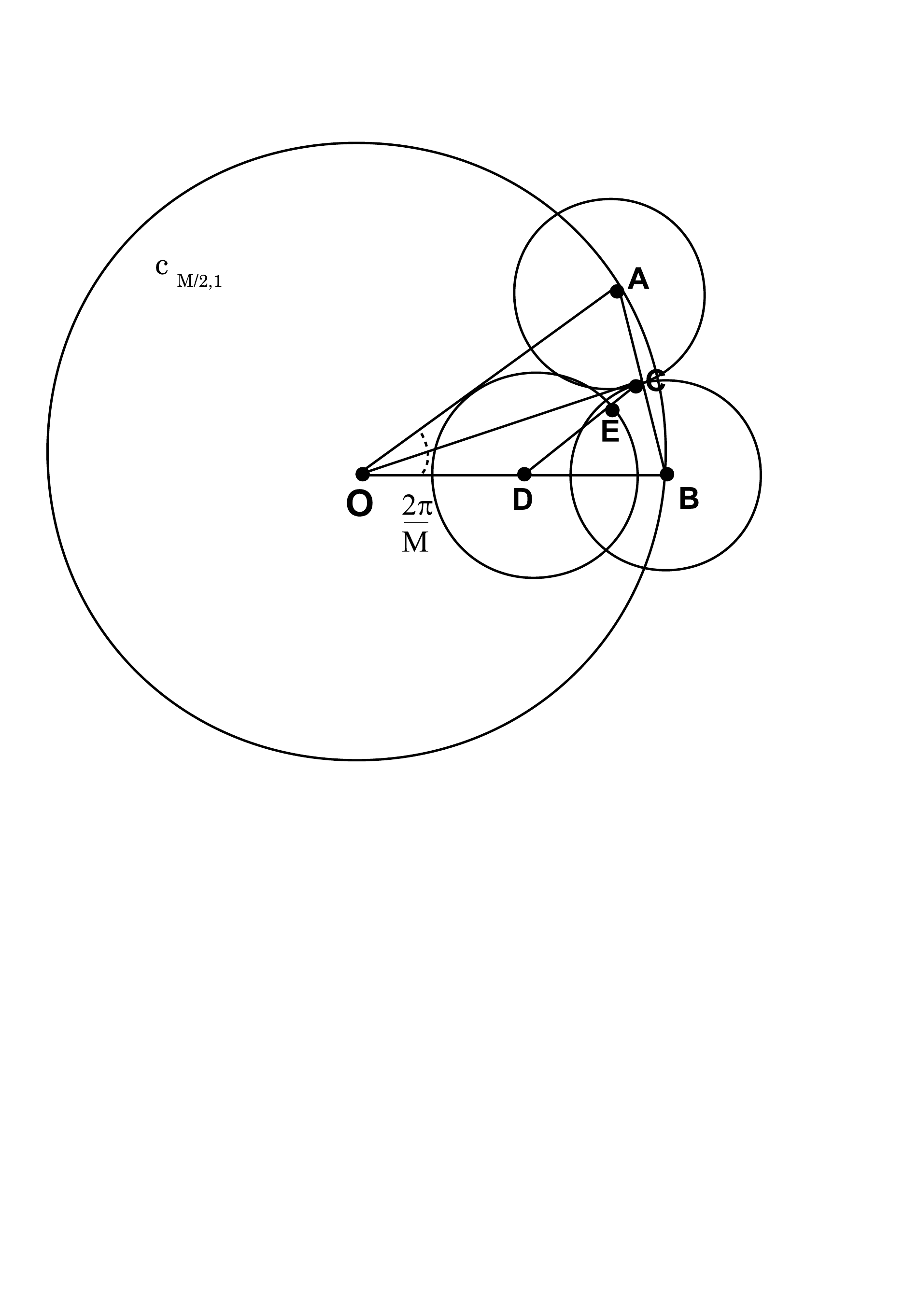}
\vspace{-7 cm}
\caption{Diagram showing the circles used in the proof of Lemma \ref{lemma_outer_two}	}	
\label{circles3}	
\end{figure}
\end{proof}
\end{lemma}

\begin{lemma}
\label{region_ext_lemma}
For $M$-PSK signal set, the region $\Gamma ^{ext}_{SF}$ is the outer envelope region formed by the origin and the $2M$ unit circles with centers at $\cot (\pi/M) e^{jk2\pi/M}$, $\mathrm{cosec} (\pi/M) e^{j(2k+1)\pi/M}, 0 \leq k \leq M-1$.
\begin{proof}
Consider the sets of circles $C_{M/2,1}$ and $C_{M/2-1,1}$.  The circles belonging to the set $C_{M/2,1}$ have centers at $1/\sin(\pi/M)e^{j(2k+1)\pi/M}=\textrm{cosec}(\pi/M)e^{j(2k+1)\pi/M}, 0 \leq k \leq M-1$ and unit radii. Similarly, the circles belonging to the set $C_{M/2-1,1}$ have centers at $\sin\left((M/2-1)\pi/M\right)/\sin(\pi/M) e^{jk2\pi/M}=\cot(\pi/M) e^{jk2\pi/M}, 0 \leq k \leq M-1$ and unit radii. Since we consider $\gamma >1$, it is enough to consider those $n,n'$ for which $c_{n,n'}$ has radius greater than or equal to one. Since for $1 \leq n,n' \leq N/2$, $\sin(\pi n/M)/\sin(\pi n'/M) \geq 1$ if and only $1 \leq n' \leq n, 1 \leq n \leq M$, it is enough to consider $n,n'$ such that $1 \leq n' \leq n, 1 \leq n \leq M$.

 Since by Lemma \ref{fully_shaded} the region between the circles $c_{M/2,1}$ and $c_{1,1}$ is fully shaded, among the sets $C_{k_1,k_2}$'s it is enough to consider those which  contain circles which lie outside  $c_{M/2,1}$. From Lemma \ref{lemma_tot_inside}, circles in those sets $C_{k_1,k_2}$, for which $k_2 =1$ alone can lie outside $c_{M/2,1}$, for $M>4$. From Lemma \ref{lemma_outer_two}, the interior of the circles which belong to the sets $C_{k_1,1}$, $k_1 \leq M/2-2$ are fully shaded. Hence the unshaded region for $\gamma > 1$ is the the exterior region of the outer envelope of the $2M$ circles which belong to the sets $C_{M/2,1}$ and $C_{M/2-1,1}$. The proof of the lemma is complete for $M>4$. For $M=4$, among all the circles $c_{n,n'}$, only $c_{2,1}$ and $c_{1,1}$ have radius greater than or equal to one. Hence Lemma \ref{lemma_tot_inside} becomes irrelevant and the proof holds for $M=4$ as well.
\end{proof}
\end{lemma}

\section*{PROOF OF THEOREM 1}
\begin{proof}
The region $\Gamma ^{ext}_{SF}$ (Region I described in Theorem 1) is given by Lemma \ref{region_ext_lemma}. If $M>4$, the $2M$ circles given in Lemma \ref{region_ext_lemma} after complex inversion become  $M$ circles with centers at $\sec (\pi/M) \tan (\pi/m) e^{j(2k+1)\pi/M}, 0 \leq k \leq M-1$ with radius $\tan^2(\pi/M)$ and $M$ circles with centers at $1/2 \tan (2\pi/M) e^{jk2\pi/M}, 0 \leq k \leq M-1$ with radius $1/2\tan(\pi/M)\tan(2\pi/M)$. If $M=4$, the $2M$ circles given in Lemma \ref{region_ext_lemma} after complex inversion become
four unit circles with centers at $\sqrt{2}e^{j\pi/2}$ and four straight lines $\gamma e^{j \theta}=\pm 0.5$, $\gamma e^{j \theta}=\pm 0.5j$. Since, by Lemma \ref{sf_int_ext}, $\Gamma ^{int}_{SF}$ is obtained by the complex inversion of $\Gamma ^{ext}_{SF}$, the region $\Gamma ^{int}_{SF}$ is given by Region II described in Theorem 1. This completes the proof.
\end{proof}

\begin{thebibliography}{} 
\bibitem{ZhLiLa}
S. Zhang, S. C. Liew and P. P. Lam, ``Hot topic: Physical-layer Network Coding'', ACM MobiCom '06, pp. 358--365, Sept. 2006.

\bibitem{KiMiTa}
S. J. Kim, P. Mitran and V. Tarokh, ``Performance Bounds for Bidirectional Coded Cooperation Protocols'', IEEE Trans. Inf. Theory, Vol. 54, pp.5235--5241, Nov. 2008.

\bibitem{PoYo}
P. Popovski and H. Yomo, ``Physical Network Coding in Two-Way Wireless Relay Channels'', IEEE ICC, Glasgow, Scotland, June 2007.

 \bibitem{KoPoTa}
T. Koike-Akino, P. Popovski and V. Tarokh, ``Optimized constellation for two-way wireless relaying with physical network coding'', IEEE Journal on selected Areas in Comm., Vol.27, pp. 773--787, June 2009.

\bibitem{KoPoTa_conv}
T. Koike-Akino, P. Popovski and V. Tarokh, ``Denoising strategy for convolutionally-coded bidirectional relaying'', IEEE ICC 2009, Dresden, Germany, June 2009.

\bibitem{HeNa}
B. Hern and K. Narayanan, ``Multilevel Coding Schemes for Compute-and-Forward'',  IEEE ISIT, St. Petersburg, Russia, July 2011.

\bibitem{Ne}
Tristan Needham, ``Visual complex analysis'', Oxford University Press, 1997.

\bibitem{Rod}
Chris A. Rodger, ``Recent Results on The Embedding of Latin Squares and Related Structures, Cycle Systems and Graph Designs.'', Le Matematiche, Vol. XLVII (1992)- Fasc. II, pp. 295-311.

\bibitem{Wa}
Ian M. Wanless, `` A Generalisation of Transversals for Latin Squares'', the electronic journal of combinatorics 9 (2002).

\bibitem{Ha}
D. E. Daykin and R. H¨aggkvist, ``Completion of sparse partial latin squares, Graph
theory and combinatorics'', 127--132, Academic Press, London, 1984

\bibitem{LS}
Vishnu Namboodiri, Vijayvaradharaj T Muralidharan and B. Sundar Rajan, ``Wireless Bidirectional Relaying and Latin Squares'', available online at arXiv: [cs.IT], Sept. 2011.

\end{thebibliography}
\end{document}